\RequirePackage{fix-cm}
\documentclass[smallextended]{svjour3}       
\smartqed 
\usepackage[T1]{fontenc}
\usepackage{appendix}
\usepackage{mathptmx}
\usepackage{amsmath}
\usepackage{graphicx}
\usepackage{float}
\usepackage[running]{lineno}
\usepackage{array}
\usepackage{longtable}
\usepackage{natbib}
\usepackage{url}
\usepackage{relsize}
\usepackage{appendix} 
\setcitestyle{aysep={}}

\newcommand*\patchAmsMathEnvironmentForLineno[1]{%
\expandafter\let\csname old#1\expandafter\endcsname\csname #1\endcsname
\expandafter\let\csname oldend#1\expandafter\endcsname\csname end#1\endcsname
\renewenvironment{#1}%
{\linenomath\csname old#1\endcsname}%
{\csname oldend#1\endcsname\endlinenomath}}%
\newcommand*\patchBothAmsMathEnvironmentsForLineno[1]{%
\patchAmsMathEnvironmentForLineno{#1}%
\patchAmsMathEnvironmentForLineno{#1*}}%
\AtBeginDocument{%
\patchBothAmsMathEnvironmentsForLineno{equation}%
\patchBothAmsMathEnvironmentsForLineno{align}%
\patchBothAmsMathEnvironmentsForLineno{flalign}%
\patchBothAmsMathEnvironmentsForLineno{alignat}%
\patchBothAmsMathEnvironmentsForLineno{gather}%
\patchBothAmsMathEnvironmentsForLineno{multline}%
}

\begin{document}

\title{$\mathcal{L}$-moments reveal the scales of momentum transport in dense canopy flows}
\titlerunning{Momentum transport in dense canopy flows}
\author{Subharthi Chowdhuri \and Olli Peltola}

\institute{Subharthi Chowdhuri \at
           Natural Resources Institute Finland (Luke)\\ 
           Latokartanonkaari 9, \\
        Helsinki, 00790, Finland\\
              \email{subharthi.chowdhuri@luke.fi}           
           \and
           Olli Peltola \at
           Natural Resources Institute Finland (Luke)\\ 
           Latokartanonkaari 9, \\
        Helsinki, 00790, Finland\\
}

\date{Received: DD Month YEAR / Accepted: DD Month YEAR}
\maketitle

\begin{abstract}
The interaction between a dense forest canopy and atmosphere is a complex fluid-dynamical problem with a wide range of practical applications, spanning from the aspects of carbon sequestration to the spread of wildfires through a forest. To delineate the eddy processes specific to canopy flows, we develop an $\mathcal{L}$-moment based event framework and apply it on a suite of observational datasets encompassing both canopy and atmospheric surface layer flows. In this framework, the turbulent fluctuations are considered as a chronicle of positive and negative events having finite lengths or time scales, whose statistical distributions are quantified through the $\mathcal{L}$ moments. $\mathcal{L}$ moments are statistically more robust than the conventional moments and have earlier been used in hydrology applications, but here we show how this concept is useful even for canopy flows. The $\mathcal{L}$-moment framework is complemented with wavelet analysis, leading to a discovery of a mixed time scale controlling the momentum exchanges between the atmosphere and the canopy air space. The origin of this mixed-scale is intimately linked to an interaction between two different eddy processes that transport momentum in the gradient and counter-gradient directions, respectively. This finding gives rise to a conceptual model of canopy turbulence that resolves a long-standing issue in canopy flows: why the integral timescale of vertical velocity increases as the heights approach the forest floor? Moreover, this model explains the intermittent nature of the wind inside a canopy despite its average being nearly zero due to canopy drag.

\end{abstract}

\section{Introduction}
\label{Intro}
Forests cover nearly 30\% of the Earth's surface and they regulate the Earth's surface temperature, emit bio-aerosol particles that act as cloud condensation nuclei, and sequester carbon from the atmosphere, thereby playing the role of a carbon sink. From a micrometeorological perspective, how the trees in a dense forest interact with the lowest layers of the atmosphere where the air is turbulent, remains a complex fluid-dynamical problem and therefore is a subject of great scrutiny \citep{raupach1981turbulence,finnigan2000turbulence,brunet2020turbulent}. Despite the complexities, this problem is important since to improve the parameterization of land-atmosphere exchanges in weather and climate models, the characteristics of turbulent transport in vegetation canopies need to be better represented \citep{harman2007simple,bonan2018modeling}. A part of this complexity arises because, unlike the atmospheric surface layer flow, the vertically distributed drag associated with the presence of a vegetation canopy and its foliage, introduces additional length scales that modulate the turbulence structure inside the canopy.

In particular, the seminal study by \citet{raupach1996coherent} introduced a mixing layer model and showed that an inflection point in the mean velocity profile at the canopy top induces Kelvin-Helmholtz (KH) instabilities that penetrate the canopy volume, which could be regarded as the canopy-scale coherent structures or the mixing layer eddies. Moreover, the flow within a canopy is obstructed by the presence of individual canopy elements, which, in turn, produce a lot of fine-scale turbulence bearing a strong resemblance with a von K\`{a}rm\`{a}n vortex shedding mechanism \citep{poggi2009flume,ghannam2015spatio}. Additionally, the attached eddies, whose sizes are proportional to the height and are commonly found in atmospheric surface layer flows, remain superposed on the canopy-scale motions. Based on this eddy structure, \citet{poggi2004effect} proposed a simple three-layer mixing-length type model for canopy flows. In this model, it was assumed that the layers deep within the canopy are only affected by the von K\`{a}rm\`{a}n eddies, the middle canopy layer has influences from both canopy scale eddies and attached eddy motions, while the upper layers of the canopy only feel the influence of the attached eddies. As one may realise, an inherent assumption in this formulation is the eddies generated at the canopy top do not exert significant influences at heights deep within the canopy. 

Later on, the studies by \citet{ghisalberti2002mixing} and \citet{chung2021turbulence} on aquatic vegetation canopies and canopies artificially generated in a laboratory, hypothesized that there exists a certain height up to which the mixing layer eddies penetrate within a canopy. \citet{chung2021turbulence} defined this height as $h-z_p$, where $h$ is the canopy height and $z_p$ is the vertical distance from the canopy floor to the height where the momentum flux falls to about 10\% of the maximum momentum flux at the canopy top. In the case of aquatic canopies, \citet{nepf2012flow} posited a length scale $\delta_e$ that gave an estimation of up to what height do the mixing layer eddies penetrate within a canopy, when the conditions are conducive for the formation of KH instabilities. This $\delta_e$ was shown to be dependent on the density of the canopy. However, for terrestrial deciduous canopies, \citet{perret2021stability} found that the eddy motions deep within a canopy (without specifying whether the canopy was dense according to the criterion of \citet{nepf2012flow}) were modulated by the canopy-scale coherent structures. A similar finding was reported by \citet{cava2022vertical}, where they demonstrated that the momentum transport events deep within the Amazon forest carried the signatures of the mixing layer eddies. They specifically highlighted the role of counter-gradient momentum transport in that regard. Recently, \citet{peltola2021physics} developed a criterion to determine whether the eddy motions occurring across the canopy depth are strongly coupled with the forest floor or not. Their criterion is contingent on how frequently the negative vertical velocity fluctuations (downdrafts) at the canopy top exceed a certain threshold, which, in turn, depends on canopy density and thermal stratification of air. Therefore, it appears that the canopy scale coherent structures or mixing layer eddies can extend their footprints deep down the canopy, although they might not actively participate in the transport of momentum.

Regarding momentum transport in canopy flows, the earlier studies found a difference between the ejection and sweep motions for heights within and above a canopy \citep{thomas2007flux,brunet2020turbulent}. In line with the quadrant nomenclature, these ejection and sweep motions were identified as the conditions when $u<0, w>0$ and $u>0, w<0$, respectively, where $u$ and $w$ are the turbulent fluctuations in the streamwise and vertical directions. These studies showed that the contributions to the momentum flux within the canopies were mostly dominated by sweeps, while above the canopies ejections played a more significant role. However, a traditional quadrant analysis does not provide any information about the time scales of these different quadrant motions. To fill that gap, \citet{chowdhuri2022scale} conducted a persistence analysis on the momentum flux events for a dense canopy flow. They considered the time scales of the four quadrants separately by simply estimating how much time do they spend in a particular quadrant state before switching to an another one. By doing so, they plotted the probability density functions (PDFs) of these time scales for the four quadrants that included the counter-gradient ones for which the momentum fluxes were positive. They found that the time scale PDFs of the ejection and sweep events remained remarkably invariant with height and displayed no changes within or above the canopy. On the contrary, the time scale PDFs of the counter-gradient events were quite sensitive to the location. For instance, these PDFs showed heavy tails within the canopy but those progressively disappeared as the heights approached the canopy top. 

These height variations in the PDFs of counter-gradient time scales were also reflected in the persistence PDFs of vertical velocity fluctuations. In this case, the time scales were defined as the times the $w$ signal spends either in a positive or in a negative state. \citet{chowdhuri2022scale} found that the persistence PDFs of $w$ signals showed a clear bulge for heights within the canopy, meaning more large-scale events populated the vertical velocity fluctuations. Therefore, the $w$ signals appeared to behave more coherently within the canopy than at the heights above it. Some previous studies had reported that the integral time scales of $w$ ($\gamma_w$) were larger within the canopy air space as compared to above, although no satisfactory explanation was provided regarding the same \citep{launiainen2007vertical,chamecki2013persistence}. In fact, from large eddy simulations, \citet{patton2016atmospheric} observed that the integral scales of vertical velocity diminished at heights above the canopy. In field-experimental datasets, this behaviour is not documented well since a more standard practice to plot the vertical profiles of $\gamma_w$ is by multiplying them with the local mean wind speed (Taylor's hypothesis), whose values itself decrease significantly within the canopy \citep{finnigan2000turbulence}. As a result, the integral length scales of vertical velocity appear small within the canopy sub layers. It is worth noting that the applicability of Taylor's hypothesis in canopy flows is questionable \citep{everard2021sweeping}. 

Despite such interesting observations, the results of \citet{chowdhuri2022scale} were qualitative in nature since the persistence PDFs were empirically determined and therefore their properties were not quantified. Due to this limitation, it was not clear what type of eddy processes give rise to the large counter-gradient events in canopy sub layers and how exactly is that different from a canonical atmospheric surface layer flow. In general, the persistence PDFs display an extended power-law regime and thus they fall into the category of heavy-tailed distributions. To quantify such heavy-tailed distributions, conventional statistical moments (such as kurtosis) do not work well since their estimates do not converge satisfactorily \citep{nair2022fundamentals}. We demonstrate this in Fig. \ref{fig:1} of this manuscript. 

As opposed to conventional moments, in this study, we deal with heavy-tailed distributions by introducing $\mathcal{L}$-moment as a statistical concept and show how it can be used in conjunction with wavelet analysis to reveal novel physical insights about canopy turbulence. $\mathcal{L}$-moments are primarily defined as linear combinations of the order statistics, which are based on the cumulative distribution functions of any stochastic signal. Although the concept of $\mathcal{L}$-moment has been used in hydrology before \citep{vogel1993moment}, but here we use it to elucidate on the physics of canopy flows. By using this framework, we seek to answer the following research questions: (1) Do the mixing layer eddies penetrate only down to a certain height within the canopy or do they reach all the way down to the forest floor? (2) Why the integral time scale of vertical velocity ($\gamma_w$) increases as the heights decrease within the canopy? (3) How these two phenomena, i.e. the penetration of mixing layer eddies and increase in $\gamma_w$, shape the transport of momentum in canopy flows?

We compare our findings between the atmospheric surface layer and canopy flows by using an extensive range of experimental datasets collected from different geographical locations spanning from the tropics to mid-latitudes. For our analysis, we restrict ourselves to near-neutral conditions, i.e. without the effects of buoyancy. The remainder of this study is organized as follows. In Section \ref{Data_method} we introduce the experimental datasets and our framework. In Section \ref{results}, the results are presented and discussed to elucidate on the canopy flow physics. Finally, in Section \ref{conclusion}, the conclusions and scopes for future research are outlined. 

\section{Dataset and methodology}
\label{Data_method}
\subsection{Dataset}
To address our research objectives, we employ two different datasets where measurements were carried out within a homogeneous and dense forest canopy. The flow over these forest canopies can be broadly categorized as roughness sublayer flows or RSL flows. The Reynolds numbers of these flows are of the order of $10^6$, assuming a boundary layer depth of 500 m. One of these datasets is the GoAmazon one, where nine level sonic anemometer measurements were available at a measurement site named Cuieiras Biological Reserve in Manaus, Brazil (3.12$^\circ$ S, 60$^\circ$ W), surrounded by the dense Amazon forest \citep{fuentes2016linking,ghannam2018scaling,chowdhuri2022scale}. The measurement heights for the GoAmazon dataset are within the range of $z/h=0.2-1.38$, where $h$ is the canopy height, approximately equal to 35 m. The leaf area index (LAI), which is defined as the total one-sided leaf area (half the total foliage area) per unit ground surface area, is estimated to be between 6.1 and 7.3 m$^{2}$ m$^{-2}$. For the Amazon forest, the vertical distribution of leaf area density showed a peak at around $z/h=0.67$, meaning that this forest had a relatively dense overstory \citep{dias2015large}. This experiment ran continuously between March 2014 and January 2015, collecting data at a 20-Hz sampling frequency. The second dataset is over Loblolly pine canopies in Duke forest (36$^\circ$ N, 78$^\circ$ W), where only one measurement height is available at $z/h=1.44$. Here, $h$ is 13 m and the sampling frequency is set at 10 Hz \citep{katul1997turbulent}. The LAI for this forest is 3.1 m$^{2}$ m$^{-2}$. For both of these RSL datasets, the data were divided into 30-min blocks and a double-coordinate rotation was applied to align the $x$-axis with the direction of the mean wind. Turbulent fluctuations in the wind components ($u$, $v$, and $w$ in the streamwise, cross-stream, and vertical directions respectively) were computed after subtracting the mean. Moreover, in our analysis, we restrict ourselves to near-neutral stratification only, satisfying the condition $|(z-d)/L|<0.5$, where, $z$ is the observation height, $d$ is the displacement height being equal to $2h/3$, and $L$ is the Obukhov length. Note that for the GoAmazon dataset, the condition $|(z-d)/L| \leq 0.5$ was imposed at and above the canopy top ($z/h=$ 1, 1.15, and 1.38). Accordingly, the $L$ values were computed from a near-constant flux profile at these three heights. This ultimately resulted in 93 and 214 blocks of 30-min runs for the GoAmazon and Duke forest datasets, respectively. While reporting the results in Section \ref{results}, an average over these ensemble of 30-min runs is performed. 

To compare the features of canopy flows with canonical atmospheric surface layer flows (ASL flows), three additional datasets were also used. One of these datasets was collected during the Surface Layer Turbulence and Environmental Science Test (SLTEST) experiment, where nine North-facing time-synchronized sonic anemometers were mounted on a 30-m mast, spaced logarithmically over an 18-fold range of heights, from 1.42 m to 25.7 m, with the sampling frequency being set at 20 Hz \citep{mcnaughton2007scaling}. This experiment was conducted at the Great Salt Lake desert in Utah, USA (40.14$^\circ$ N, 113.5$^\circ$ W), where the surface conditions were aerodynamically smooth with roughness lengths of the order of millimeters. The other two datasets were obtained during an experimental campaign in India. This experiment is known as the Cloud Aerosol Interaction and Precipitation Enhancement Experiment (CAIPEEX), during which two micrometeorological towers of 20-m height were set up over Mahbubnagar (16.75$^\circ$ N, 78$^\circ$ E) and Varanasi (25.32$^\circ$ N, 83$^\circ$ E) regions. On these towers, only at a single measurement height, the high-frequency observations of the three velocity components were sampled at a frequency of 10-Hz \citep{chowdhuri2019evaluation}. The site conditions were representative of a typical grassland with roughness lengths of the order of centimeters, an order of magnitude higher than the SLTEST experiment. Henceforth, we refer to these grassland sites as CPX1 and CPX2, respectively. The measurement heights for these experiments were 5 m and 6 m, respectively. Similar to canopy flows, the results reported in Section \ref{results} are averaged over an ensemble of near-neutral runs of 30-min duration each. The near-neutral runs are identified as those satisfying the condition $|z/L|<0.5$, which resulted in 19 and 170 blocks of 30-min runs for the SLTEST and CAIPEEX datasets, respectively. Despite the number of near-neutral runs being less for the SLTEST datasets, the second-order turbulence statistics ($u$, $w$ spectra and $u$-$w$ cospectra) were well-converged for both $u$ and $w$ signals.

\subsection{$\mathcal{L}$-moment analysis}
The conventional statistical moments are defined as, 
\begin{equation}
\overline{X^{m}}=\int_{-\infty}^{+\infty}X^{m}P(X)dx,
\label{sm}
\end{equation}
where $X$ is a stochastic signal, $P(X)$ is its probability density function (PDF), and $m$ is the moment order (e.g., $m=2$ represents variance). The sample estimates of such conventional moments, such as kurtosis ($\mathcal{K}(X)$), are computed in standardised format as, 
\begin{equation}
\mathcal{K}(X)=\overline {{(\frac{X-\overline{X}}{\sigma_{X}})}^4},
\label{sm}
\end{equation}
where $\overline{X}$ is the sample mean and $\sigma_{X}$ is the standard deviation. On the other hand, the $\mathcal{L}$ moments are defined with respect to the cumulative distribution functions (CDFs), instead of the PDFs. As shown by \citet{vogel1993moment}, the first four $\mathcal{L}$ moments are theoretically defined as,
\begin{equation}
  \begin{aligned} 
\mathcal{L}_1 & =\beta_0 \\
\mathcal{L}_2 & =2\beta_1-\beta_0\\
\mathcal{L}_3 & =6\beta_2-6\beta_1+\beta_0\\
\mathcal{L}_4 & =20\beta_3-30\beta_2+12\beta_1-\beta_0,
  \end{aligned}
\end{equation}
where $\beta_r=\mathcal{E}[X{(F_{X})}^r]$, with $X$ being a stochastic signal, $F_{X}$ is the cumulative distribution function of $X$, and $\mathcal{E}$ is the expected value. For all practical purposes, these four $\mathcal{L}$ moments can be computed from a stochastic sample $X$ of size $N$ as \citep{wang1996direct},
\begin{equation}
  \begin{aligned} 
\mathcal{L}_1 & =\frac{1}{\binom{N}{1}}\sum_{i=1}^{N}X(i) \\
\mathcal{L}_2 & =\frac{1}{2\binom{N}{2}}\sum_{i=1}^{N}\big[\binom{i-1}{1}-\binom{N-i}{1}\big]X(i) \\
\mathcal{L}_3 & =\frac{1}{3\binom{N}{3}}\sum_{i=1}^{N}\big[\binom{i-1}{2}-2\binom{i-1}{1}\binom{N-i}{1}+\binom{N-i}{2}\big]X(i) \\
\mathcal{L}_4 & =\frac{1}{4\binom{N}{4}}\sum_{i=1}^{N}\big[\binom{i-1}{3}-3\binom{i-1}{2}\binom{N-i}{1}+3\binom{i-1}{2}\binom{N-i}{2}+\binom{N-i}{3}\big]X(i),
  \end{aligned}
  \label{Lk}
\end{equation}
where $\binom{N}{k}$ is the binomial coefficient. As one can see, $\mathcal{L}_1$ is simply the sample average, $\mathcal{L}_2$ is a measure of scale, while the higher order $\mathcal{L}$-moments, $\mathcal{L}_3$ and $\mathcal{L}_4$, are used to define the measures of skewness and kurtosis. A MatLab code is available to compute these $\mathcal{L}$ moments for a stochastic sample \citep{kob2024}. \citet{hosking1990moments} introduced non dimensional $\mathcal{L}$-moment ratios and defined $\mathcal{L}$-skewness ($\mathcal{L}_s$) and $\mathcal{L}$-kurtosis ($\mathcal{L}_k$) as, 
$\mathcal{L}_s=\mathcal{L}_3/\mathcal{L}_2$ and $\mathcal{L}_k=\mathcal{L}_4/\mathcal{L}_2$, respectively. Unlike the conventional statistical moments, $\mathcal{L}$-moment ratios are bounded so that $\mathcal{L}_s$ lies within ($-1,1$), and $\mathcal{L}_k$ within ($1.25\mathcal{L}_s^{2}-0.25,1$). Such bounds are considered an advantage because it is easier to interpret bounded measures than the conventional skewness and kurtosis moments, which can take arbitrarily large values. One another advantage of $\mathcal{L}$-moments is these are particularly well suited for the analysis of heavy-tailed distributions because, unlike the conventional moments, they are finite for all distributions that have finite means. Even for distributions with tails so heavy that the mean is infinite, $\mathcal{L}$-moments provide effective tools for statistical inference \citep{hosking2007some}. As a point of reference, $\mathcal{L}_s$ and $\mathcal{L}_k$ for Gaussian distribution are 0 and 0.1226, respectively.

\begin{figure*}[h]
\centering
\includegraphics[width=0.5\textwidth]{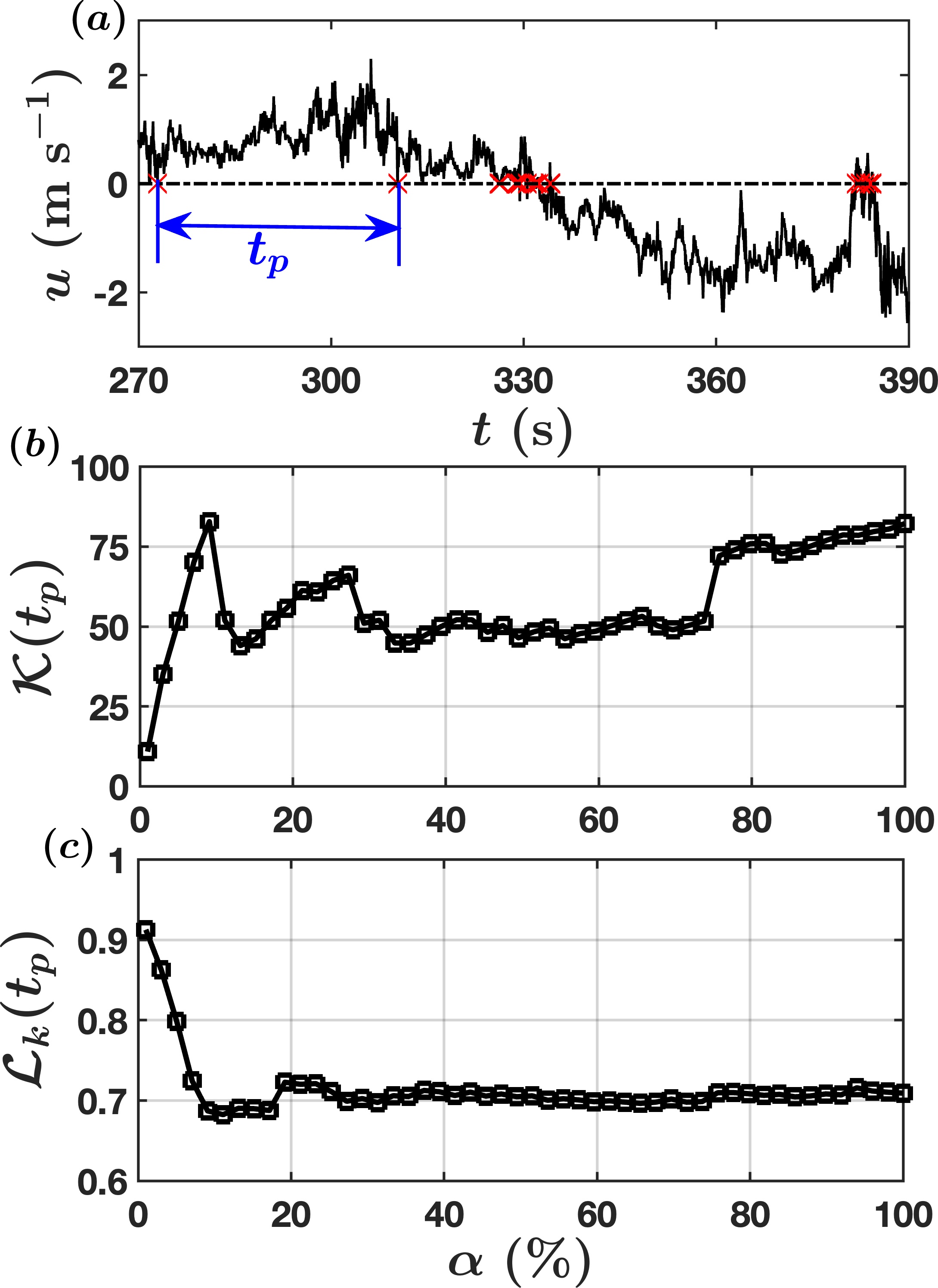}
  \caption{(a) A sample time series of the streamwise velocity fluctuation ($u$) is shown, where $t_{p}$ indicates the zero-crossing or persistence time scale while the red crosses represent the zero-crossing points. (b) The conventional kurtosis of $t_{p}$ values ($\mathcal{K}(t_p)$) are shown by changing the percentage of zero-crossing samples ($\alpha$) used for computation. (c) The same information is shown but for $\mathcal{L}$-kurtosis of $t_{p}$ values ($\mathcal{L}_{k}(t_p)$).}
\label{fig:1}
\end{figure*}

The heavy-tail distributions are fairly common across various disciplines. For example, \citet{newman2005power} argues that the distributions of sizes of cities, earthquakes, forest fires, solar flares, moon craters, and personal fortunes all exhibit power-law behavior and thus fall into the category of heavy-tailed distribution. In turbulent flows, the heavy-tailed distributions appear when one considers the PDFs of the zero-crossing time scales or in other words, the persistence time scales. For instance, let us consider a sample turbulent signal of velocity fluctuations ($u$ signal) from an atmospheric surface layer flow (Fig. 1a). In Fig. 1a, the zero-crossing points are denoted by the red crosses and the time between two successive crossings is denoted by $t_p$. The $t_p$ can also be interpreted as the time scales of the positive and negative events, and can be written as $t_p=N_p/f_s$, where $N_p$ is the event length and $f_s$ is the sampling frequency. Note that $N_p$ and $t_p$ can be used interchangeably although unlike $t_p$, $N_p$ is a discreet quantity. Notwithstanding these subtleties, if one plots the PDFs of $t_p$ with a standard normalization, $(t_p-\overline{t_p})/\sigma_{t_p}$, where $\overline{t_p}$ and $\sigma_{t_p}$ are the mean and standard deviation, and compare them with an equivalent Gaussian distribution, it can be clearly seen that these PDFs display quite a heavy tail (see Fig. S1 in the Supplementary Material). Such heavy tails are associated with large $t_p$ values, often exceeding the integral scales and thus represent the coherent structures in atmospheric flows. These heavy tails arise because the PDFs of $t_p$ display an extended power-law behaviour \citep{chamecki2013persistence,chowdhuri2020persistence}. 

An important statistical quantity to characterize these heavy tails is the kurtosis or the fourth-order moment. Typically, the computations of such higher-order moments require large sample sizes to ensure statistical convergence. This is because the conventional kurtosis moments involve the fourth power of a stochastic signal, thereby giving more weightage to the extremes that are poorly sampled when the sample space is small. This is particularly relevant for a stochastic signal such as $t_{p}$, for which the sample size is rather limited. For instance, considering a 30-min run sampled at a frequency of 20 Hz, the number of $t_{p}$ samples would hardly be of the order of 1000. Therefore, to highlight this converge issue, in Fig. \ref{fig:1}b, we show the kurtosis of $t_p$, $\mathcal{K}(t_{p})$ (see Eq. \ref{sm}), after artificially reducing the number of $t_p$ samples. This reduction is shown as a percentage of the original sample length and denoted as $\alpha$. One can notice that the $\mathcal{K}(t_{p})$ values suffer from convergence as they do not attain a plateau even when $\alpha$ reaches nearly 100\%. On the other hand, if one plots $\mathcal{L}_{k}(t_{p})$, computed from Eq. \ref{Lk}, a quite fast convergence is achieved as the values become stable even at an $\alpha$ as small as 20\% (Fig. \ref{fig:1}c). This establishes the superiority of the $\mathcal{L}$-kurtosis moments to quantify the heavy tails of the event length or persistence PDFs.  

In general, the distribution of zero-crossing time scales is a non-trivial problem and poses serious theoretical challenges to derive them from the first principles \citep{majumdar1999persistence}. Therefore, a bounded statistical measure such as $\mathcal{L}$-kurtosis to quantify these non-trivial distributions is of significant value. To associate these $\mathcal{L}$-kurtosis moments further with the underlying turbulence physics, we generate a scale-wise description of $\mathcal{L}_{k}(N_p)$ values. For convenience purposes, we use the event lengths $N_p$. The scale-wise description is achieved by computing the $\mathcal{L}_{k}(N_p)$ values of the velocity increment signals (for instance, $\Delta u$), which is defined as $u(t+\Delta t)-u(t)$, where $\Delta t$ is the time lag. These time lags represent the eddy time scales in the flow. Thus, if one plots the $\mathcal{L}_{k}(N_p)$ values against $\Delta t$, such graphs would reveal the role of the eddy motions towards generating the large events in a turbulent flow field. Henceforth, the scale-wise $\mathcal{L}_{k}(N_p)$ values will be denoted simply as, $\mathcal{L}^{\Delta u}_{k}$ and the $\mathcal{L}_{k}(N_p)$ values corresponding to the full signals will be written as, $\mathcal{L}^{u}_{k}$. 

\subsection{Wavelet analysis}
In addition to the $\mathcal{L}$-moment analysis, we carried out a wavelet analysis on the $u$ and $w$ signals to compute their cospectra and coherence spectra. As will be shown later (see Section \ref{results}), regarding canopy flows, the wavelet analysis provides some interesting observations, which are further explained through the $\mathcal{L}$-moment analysis. Therefore, these two different methods complement each other quite nicely. 

For the wavelet analysis, we used ‘Morlet’ as the mother wavelet due to its localization properties. The wavelet scales were converted to equivalent Fourier frequencies ($f$), using the relationship provided by \citet{torrence1998practical}. The premultiplied global wavelet cospectrum between $u$ and $w$ ($fS_{uw}(f)$) was defined as,
\begin{equation}
fS_{uw}(f)=\frac{\delta t}{N}\sum_{n=1}^{N}\mathcal{R}\big[\frac{W^{u}_{n}(s_j) \overline{W^{w}_{n}(s_j)}}{s_j}\big],
\label{wavelet}
\end{equation}
where $s_j$ are the wavelet scales, $W^{u}_{n}(s_j)$ is the wavelet coefficient for the $u$ signal at scale $s_j$, $\overline{W^{w}_{n}(s_j)}$ is the complex conjugate of the wavelet coefficient for the $w$ signal at scale $s_j$, $\mathcal{R}$ is the real component of a complex number, $N$ is the number of data points in a time series, $J$ is the total number of scales, and $\delta t$ is the sampling period, which is equal to $1/f_s$ with $f_s$ being the sampling frequency. Equation \ref{wavelet} can also be used to compute the spectra if instead of two different signals similar ones are used.

The coherence spectrum, on the other hand, is defined as,
\begin{equation}
\Gamma_{u,w}^2(f)=\frac{1}{N}\sum_{n=1}^{N}\frac{|\mathcal{R}(W^{uw})+\mathcal{C}(W^{uw})|^2}{W^{uu}W^{ww}},
\label{coherence}
\end{equation}
where $\Gamma_{u,w}^2(f)$ is the squared coherence between $u$ and $w$, $W^{xy}=W^{x}_{n}(s_j) \overline{W^{y}_{n}(s_j)}$ with $x$ and $y$ being equal to either $u$ or $w$, and $\mathcal{R}$ and $\mathcal{C}$ are the real and imaginary components of a complex number. Henceforth, $f$ will be removed from $\Gamma_{u,w}^2(f)$ while showing the results. The same procedure can be repeated to compute the squared coherence between any two signals. 

We also conducted a wavelet cross-scalogram analysis on the $u$ and $w$ signals.  Wavelet cross-scalograms include time information in addition to coherence and phase. We implemented this based on the MatLab code provided in \citet{grinsted2004application} with default choices such as: a mother wavelet of ‘Morlet’; a scale resolution of 10 scales per octave; and the addition of zeros at the end to increase the length of the time series to the nearest power of 2. This artificial addition of zeros added a ‘cone of influence’, within which the results were not interpreted as they are impacted by the edge effects.

\section{Results and discussion}
\label{results}
We begin by showing how the concept of $\mathcal{L}$-kurtosis can be effectively applied to gain valuable insights into RSL flows and highlight their differences with respect to ASL flows. In our analysis we specifically focus on the streamwise and vertical velocity fluctuations ($u$ and $w$), since these two signals together explain the vertical momentum transport ($uw$) between the upper atmosphere and the canopy air space. These results are complemented with wavelet analysis, leading to a discovery of a mixed time scale that controls the momentum exchanges in RSL flows. The origin of this mixed time scale is dissected further by first conditionally sampling the momentum fluxes into its gradient ($uw<0$) and counter-gradient ($uw>0$) components and later carrying out an $\mathcal{L}$-kurtosis analysis on their time scales. It is shown that this mixed time scale represents an interaction between two different eddy processes that transport momentum in the gradient and counter-gradient directions, respectively. We end our discussion by proposing a conceptual model of canopy turbulence that explains why the integral scales of vertical velocity decrease with height in canopy sub-layer.

\subsection{Contrast between ASL and RSL flows}
\label{contrast}
\begin{figure*}[h]
\centering
\includegraphics[width=1\textwidth]{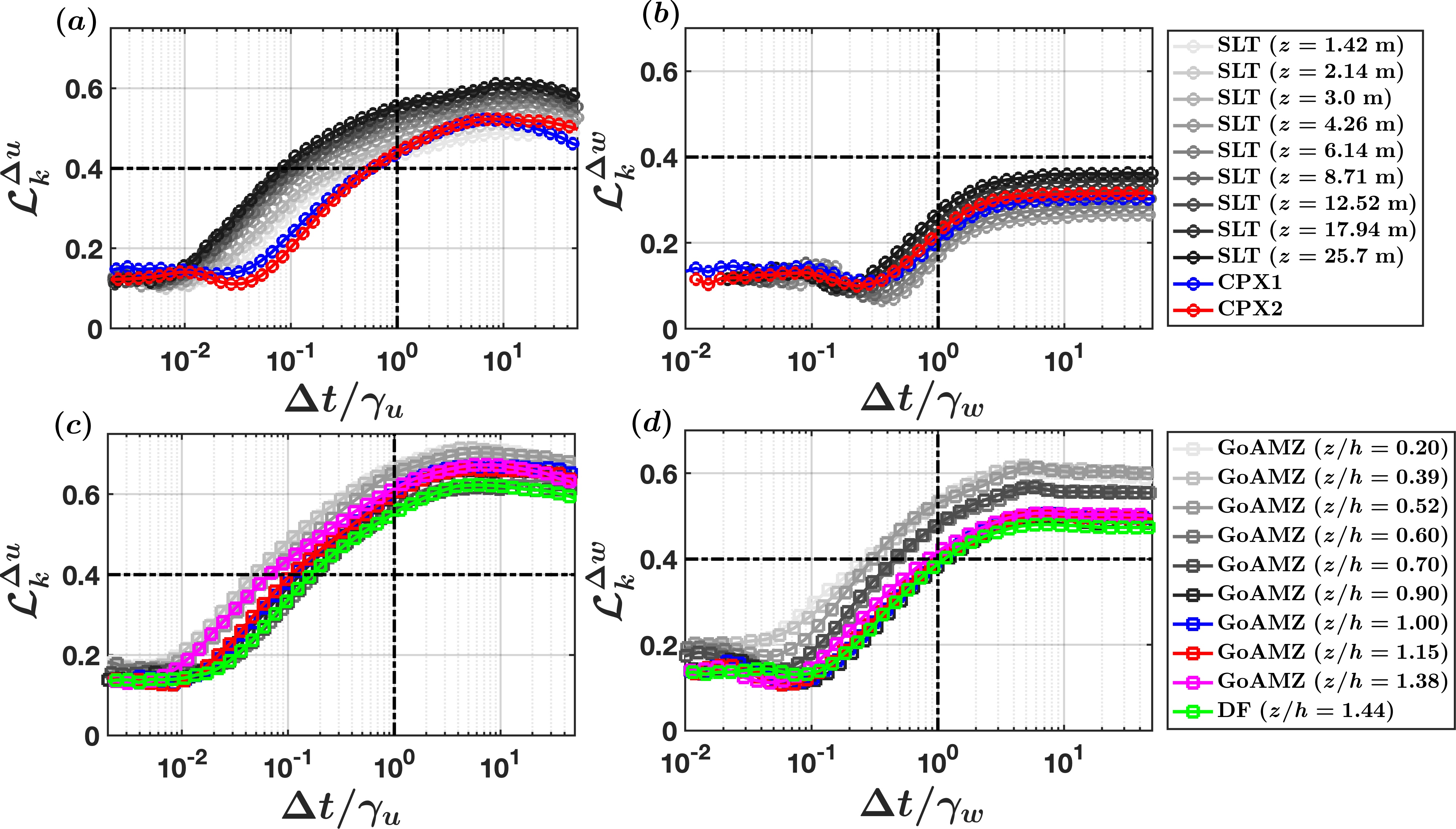}
  \caption{The $\mathcal{L}$-kurtosis values of the event lengths are shown for the (a) $\Delta u$ ($\mathcal{L}_{k}^{\Delta u}$) and (b) $\Delta w$ ($\mathcal{L}_{k}^{\Delta w}$) signals, corresponding to ASL flows. Here, $\Delta u$ and $\Delta w$ represent the velocity increments at a time lag $\Delta t$. The time lags are normalized by the integral time scales of $u$ and $w$, respectively ($\gamma_{u}$ and $\gamma_{w}$). (c,d) In the bottom panels, the same information are shown but for the RSL flows. The legends at the right-hand side describe the different ASL and RSL datasets used in the present analysis. For brevity, SLTEST, GoAmazon, and Duke Forest datasets are labeled as SLT, GoAMZ, and DF, respectively. The horizontal dash-dotted lines denote the critical $\mathcal{L}$-kurtosis value of 0.4, beyond which the PDF-based moments cease to exist.}
\label{fig:2}
\end{figure*}

Figure \ref{fig:2} depicts the scale-wise evolution of the $\mathcal{L}$-kurtosis ($\mathcal{L}_{k}$) values of the event lengths, corresponding to the $\Delta u$ ($\mathcal{L}_{k}^{\Delta u}$) and $\Delta w$ ($\mathcal{L}_{k}^{\Delta w}$) signals. Here, $\Delta x$ (where $x=u,w$) are the velocity increments, defined as $\Delta x=x(t+\Delta t)-x(t)$, where $\Delta t$ is the prescribed time lag. The time lags are normalized by the integral time scales of $u$ ($\gamma_{u}$) or $w$ ($\gamma_{w}$), depending on the signal types. As per the standard practice, the integral scales are computed by integrating the autocorrelation functions of the velocity signals up to their first zero-crossings \citep{chamecki2013persistence}. The curves shown in Fig. \ref{fig:2} are ensemble averaged over all the near-neutral runs, and their run-to-run variations, expressed as one standard deviation from the ensemble mean, are found to be small in all the cases. The associated error-bar plots, corresponding to RSL flows, are shown as an example in Fig. S2 of the Supplementary Material. Throughout this study we use integral time scales rather than converting them to length scales through Taylor's hypothesis, since its applicability remains questionable in RSL flows.

The upper two panels in Fig. \ref{fig:2} show the plots from the ASL flows (Figs. \ref{fig:2}a--b), while the bottom two panels show the same for the RSL flows (Figs. \ref{fig:2}c--d). The multiple heights from the SLTEST and GoAmazon datasets are color-coded in gray shades with their intensities increasing as the heights increase. Regardless of the signal types, the $\mathcal{L}_{k}^{\Delta x}$ values monotonically increase with increasing time-lags and eventually attain a plateau at scales comparable to the integral scales. As shown in Appendix \ref{app_A}, the $\mathcal{L}_{k}^{\Delta x}$ values associated with these plateaus are equal to $\mathcal{L}_{k}^{x}$, i.e. the values obtained from the event lengths ($N_p$) of the full signals. This signifies that the long-duration events in the velocity signals are formed when the larger-scales in the flow are accounted for. This mechanism is analogous to how the very-large-scale motions (VLSMs) are formed in the log-layers of wall-bounded flows, where it is argued that these VLSMs or long-duration motions are created by an accumulation of coherent structures in the flow \citep{deshpande2023evidence}. Coming back to Fig. \ref{fig:2}, the attainment of a plateau at scales of the order of $\gamma_{x}$ indicates that the heavy tails of the event length PDFs of $u$ or $w$ signals are clearly associated with the large-scale coherent structures being passed over the measurement location. Despite such similarities, clear differences appear between the ASL and RSL flows when the $u$ and $w$ signals are separately looked at. 

For instance, one can notice nearly no discernible differences between the ASL and RSL flows when $\mathcal{L}_{k}^{\Delta u}$ curves are considered. For both of these flows, $\mathcal{L}_{k}^{\Delta u}$ curves cross 0.4 after a threshold time scale. The significance of $\mathcal{L}_{k}^{\Delta u} \geq 0.4$ comes from the literature of $\mathcal{L}$-moments, where it is postulated that the $\mathcal{L}$-kurtosis exceeding 0.4 indicates that the conventional PDF-based moments do not exist for a stochastic signal \citep{nair2022fundamentals}. This is interesting because the mean of the zero-crossing rate of a turbulent signal is associated with the dissipation rate of the turbulence kinetic energy \citep{sreenivasan1983zero}. For future studies, it remains to be seen whether the $\mathcal{L}$-kurtosis values of the event lengths can be used to correct the dissipation rate estimates of the turbulence kinetic energy. 

\begin{figure*}[h]
\centering
\includegraphics[width=1\textwidth]{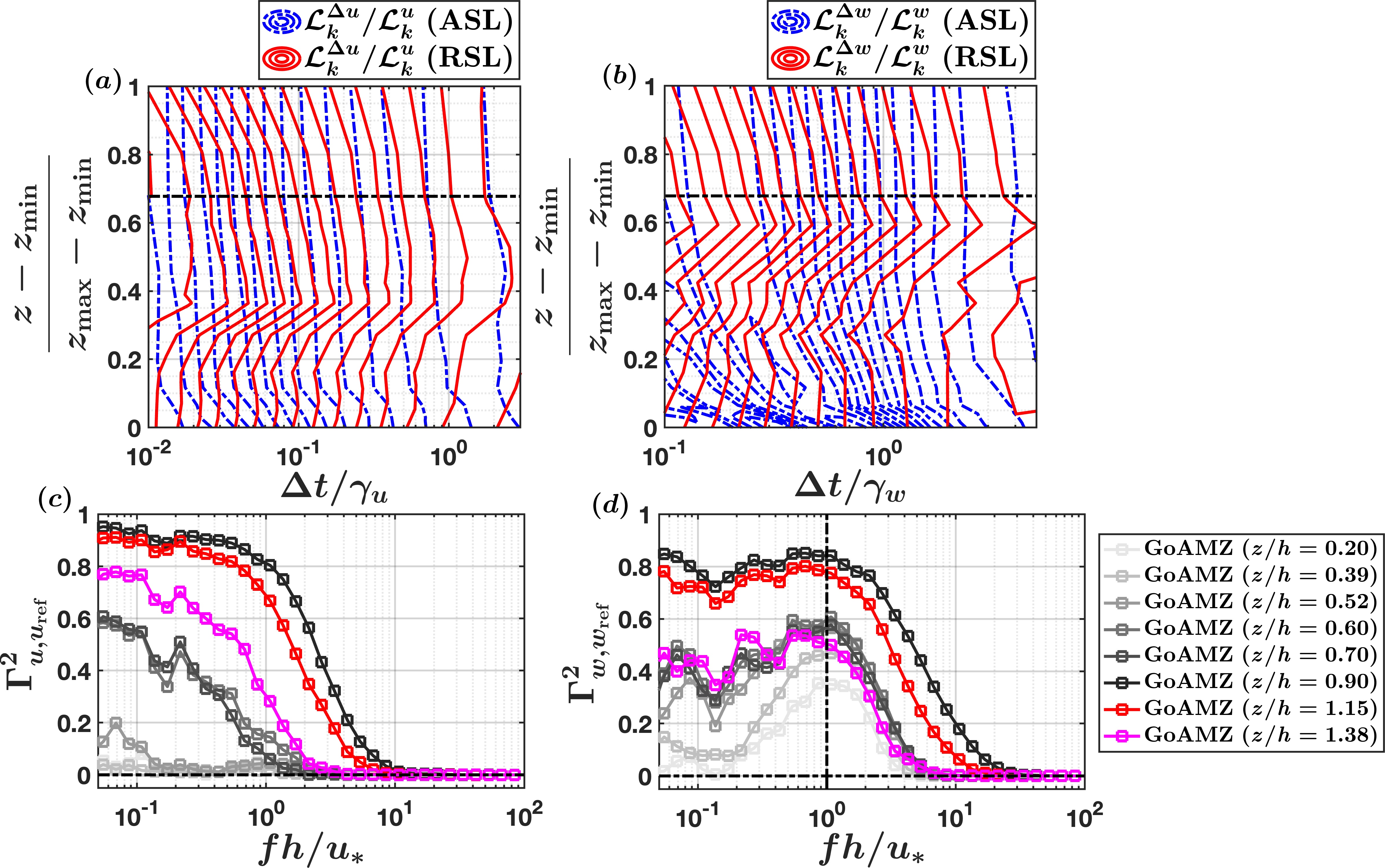}
  \caption{The time-height contour plots of (a) $\mathcal{L}_{k}^{\Delta u}$ and (b) $\mathcal{L}_{k}^{\Delta w}$ are shown to compare the features between the ASL and RSL flows. For this purpose, the SLTEST and GoAmazon datasets are used to represent the ASL and RSL flows, respectively. Note that the $\mathcal{L}_{k}^{\Delta x}$ ($x=u,w$) are normalized with the $\mathcal{L}_{k}$ values of the event time scales as obtained from the full-signals ($\mathcal{L}_{k}^{x}$). The heights ($z$) are scaled as $(z-z_{\rm min})/(z_{\rm max}-z_{\rm min})$. The contours are spaced in the increments of 0.05 for both ASL and RSL datasets. Corresponding to RSL flows, the wavelet-based coherence spectra are shown for (c) $u$ and (d) $w$ signals, where $\Gamma^2_{x,x_{\rm ref}}$ represents the squared coherence between the $x$ and $x_{\rm ref}$ signals. Here, $x_{\rm ref}$ indicates the reference signals exactly at the canopy top ($z/h=1$) while the $x$ signals are from heights other than that. The frequencies are normalized by the canopy time scale ($fh/u_{*}$) and the vertical dash-dotted line in (d) denotes the position $fh/u_{*}=1$. The $h$ denotes the canopy height while $u_{*}$ is the friction velocity at the canopy top. The different heights from the GoAmazon dataset are shown in the legend.}
\label{fig:3}
\end{figure*}

On the other hand, the $\mathcal{L}_{k}^{\Delta w}$ curves differ significantly between the ASL and RSL flows. From Fig. \ref{fig:2}b, it is quite apparent that the $\mathcal{L}_{k}^{\Delta w}$ values remain considerably smaller than 0.4 at all scales of the flow. Conversely, for the RSL flows, similar to $\mathcal{L}_{k}^{\Delta u}$, $\mathcal{L}_{k}^{\Delta w}$ values exceed 0.4. This indicates that the PDFs of the event lengths are heavier than the ASL flows, thereby signifying more coherence in $w$ signals of RSL flows as compared to a canonical atmospheric surface layer. Moreover, the $\mathcal{L}_{k}^{\Delta w}$ curves of RSL flows separate from one another as the heights approach the canopy top indicative that long events are more prevalent in $w$ time series close to the forest floor than higher up. These discrepancies get further highlighted when one plots the time-height contours of $\mathcal{L}_{k}^{\Delta x}$ values (Figs. \ref{fig:3}a--b). 

In Figs. \ref{fig:3}a--b the information presented in Fig. \ref{fig:2} are shown in the form of contour plots. In comparison to line plots, the contour representation of $\mathcal{L}_{k}^{\Delta x}$ values is more beneficial to differentiate better between the ASL and RSL flow features. The $x$ axes of Figs. \ref{fig:3}a--b denote the normalized time lags, while the $y$ axes represent the standardized heights. For this purpose, only the SLTEST and GoAmazon datasets are used to represent the ASL and RSL flows, respectively, since they contain multi-level measurements. Given their different height ranges, the $z$ values are scaled as $(z-z_{\rm min})/(z_{\rm max}-z_{\rm min})$. In this scaled coordinate system, the canopy top appears at a value of around 0.75, which is shown as a dash-dotted horizontal line in Figs. \ref{fig:3}a--b. Moreover, as the ranges of $\mathcal{L}_{k}^{\Delta x}$ values also differ between the two datasets (Fig. \ref{fig:2}), they are normalized by their full signal values and therefore remain bounded between 0 to 1 (see Fig. \ref{fig:8} in Appendix \ref{app_A}). The contours associated with the ASL and RSL datasets are identified by different colours and line types, blue (dash-dotted lines) and red (solid lines), respectively.

After carrying out these exercises, one can notice the striking similarities between the blue and red contours of the $u$ signals in Fig. \ref{fig:3}a. In fact, except the lowest few levels, the contours of $\mathcal{L}_{k}^{\Delta u}/\mathcal{L}_{k}^{u}$ agree remarkably well between the ASL and RSL flows at all time scales. Since the streamwise velocity fluctuations are more influenced by the eddies whose sizes scale with the atmospheric boundary layer (ABL) depth, this finding underscores the importance of the ABL-scale motions for both ASL and RSL flows \citep{dupont2022influence}. On the contrary, the blue and red contour lines of $\mathcal{L}_{k}^{\Delta w}/\mathcal{L}_{k}^{w}$ clearly differ from one another at heights $(z-z_{\rm min})/(z_{\rm max}-z_{\rm min})<0.75$ (within the canopy), albeit they match well for heights above the canopy. Similar conclusions can be drawn, if instead of the $\mathcal{L}$-kurtosis values, the $\mathcal{L}$-skewness was used to the quantify the heavy tails of the event length PDFs (see Fig. \ref{fig:9} in Appendix \ref{app_B}). Therefore, these robust findings indicate that the dynamics of the vertical velocity fluctuations encode the effects of canopy-scale eddies in RSL flows, which is a specific feature of canopy turbulence only. More importantly, the influences of the canopy-scale eddies are not restricted up to any specific height but they extend down to heights approaching the forest floor. We next perform a wavelet analysis to lend more credence to this hypothesis. 

To determine whether the canopy-scale eddies indeed act differently on the streamwise and vertical velocity fluctuations, we carried out a wavelet coherence analysis on the GoAmazon dataset. In this analysis, the squared wavelet coherence ($\Gamma^2_{x,x_{\rm ref}}$, $x=u,w$)  was computed between the two signals, where the reference signal ($x_{\rm ref}$) was located exactly at the canopy top ($z/h=1$) while the other ones ($x$) were sampled from rest of the heights. Physically, the squared coherence estimates indicate how strongly the two signals are linearly correlated with each other at each frequency. For this purpose, the 'Morlet' wavelet was chosen as the mother wavelet and the frequencies ($f$, converted from wavelet scales) were scaled with the canopy time scale $h/u_{*}$. The procedure to compute this coherence is same as shown in Eq. \ref{coherence}. Note that the quantity $h/u_{*}$ is height-invariant and therefore can be considered as a global time scale \citep{brunet2020turbulent}. 

By comparing the scale-wise behaviour of $\Gamma^2_{u,u_{\rm ref}}$ with $\Gamma^2_{w,w_{\rm ref}}$, it is conspicuous that the $\Gamma^2_{w,w_{\rm ref}}$ curves attain a clear peak at a time scale of $fh/u_{*}=1$ for all the $z$ values (Fig. \ref{fig:3}d). On the other hand, the $\Gamma^2_{u,u_{\rm ref}}$ curves remain close to zero for the lowest three heights of the GoAmazon dataset (Fig. \ref{fig:3}c). This observation raises an important point. If the $u$ signals are considered solely, then one might interpret that the canopy-scale eddies only exert their influences up to a certain height since the lowest three levels appear to be disconnected with the turbulent processes occurring at the canopy top. However, the same conclusion does not hold in the case of $w$ signals, for which the influences of the canopy eddies can even be felt quite strongly at those lowest three heights. In fact, similar outcomes are obtained, if instead of a coherence analysis, a lead-lag correlation analysis were carried out with the reference signals being located at the canopy top. Thereby, these findings can be considered to be robust. Since the vertical velocity fluctuations are the dominant carriers of momentum across the atmosphere and the canopy air space, we next investigate the impact of these canopy-scale eddies on the momentum transport in RSL flows. 

\subsection{Momentum transport in RSL flows}
\label{momentum flux}
\begin{figure*}[h]
\centering
\includegraphics[width=1\textwidth]{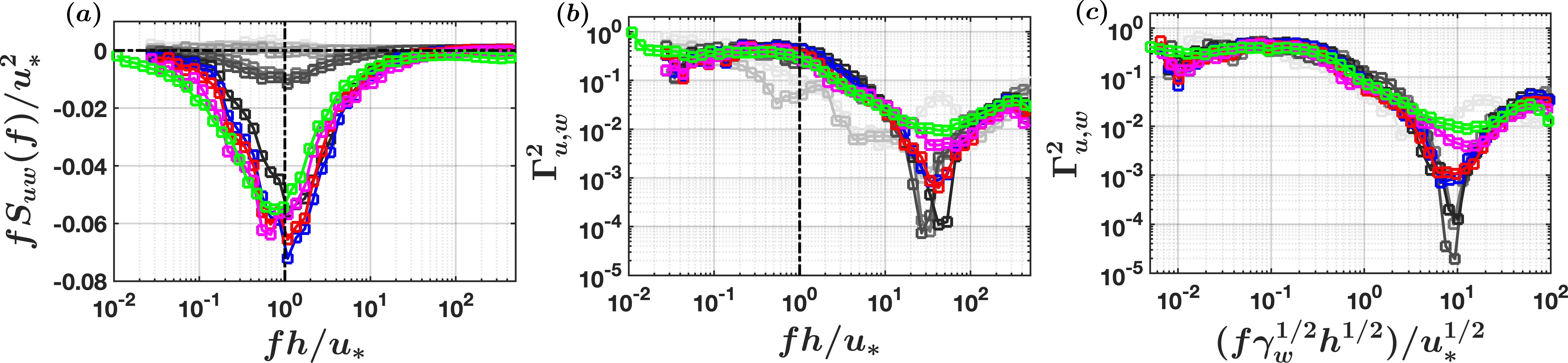}
  \caption{For the RSL flows, the wavelet-based (a) premultiplied cospectra ($fS_{uw}(f)$) and (b) coherence spectra ($\Gamma^2_{u,w}$) are shown between the $u$ and $w$ signals. The cospectral amplitudes are normalized by $u_{*}^2$ and the frequencies ($f$) are normalized by the canopy time scale. The vertical dash-dotted lines in (a) and (b) denote the position $fh/u_{*}=1$. (c) The same coherence spectra between $u$ and $w$ signals are shown but the frequencies are normalized with a mixed time scale, ${(\gamma_{w}h/u_{*})}^{1/2}$, where $\gamma_{w}$ is the integral time scale of the $w$ signal. Different colors represent the two different RSL datasets as shown in the legend of Fig. \ref{fig:2}c--d.}
\label{fig:4}
\end{figure*}

Figure \ref{fig:4}a shows the premultiplied wavelet cospectra ($fS_{uw}(f)$) between the $u$ and $w$ signals for the RSL flows. As usual, the frequencies are normalized with the canopy time scale and the cospectral amplitudes are scaled by the square of the friction velocity at the canopy top. The cospectral amplitudes decrease rapidly towards zero as measurement height decreases below $z=h$, since most of the momentum are absorbed by the upper parts of the canopy. For heights above the canopy, the cospectral amplitudes show a clear peak at a scaled frequency commensurate with $fh/u_{*}=1$ albeit slight height dependence is observable. As opposed to the cospectral amplitudes, if one investigates the squared coherence between the $u$ and $w$ signals, the $\Gamma^2_{u,w}$ values remain considerably larger even for heights deep within the canopy (Fig. \ref{fig:4}b). Especially, these high $\Gamma^2_{u,w}$ values are found at frequencies $fh/u_{*} \leq 1$. Therefore, this finding indicates that the canopy-scale eddies insert their influences at heights deep within the canopy, although they do not transport any momentum in an averaged sense. However, it is not solely the canopy-scale eddies that dictate this high coherence. In fact, these coherence curves can be collapsed reasonably well if a mixed time scale is used in place of the canopy time scale (Fig. \ref{fig:4}c). This mixed time scale, ${(\gamma_{w}h/u_{*})}^{1/2}$, is a geometric mean between the integral time scale of the vertical velocity fluctuations and the canopy time scale. 

In the context of turbulence literature, the mixed scale has been observed before but no satisfactory explanation exists on why such scaling appears while collapsing the turbulence statistics \citep{buschmann2009near,gadeffects}. It is hypothesized that this scaling represents an interaction between two different eddy processes occurring at two time scales that constitute the mixed scale \citep{mcnaughton2007scaling}. With respect to ASL flows, the evidence of mixed scale has earlier been reported in the context of both convective \citep{chowdhuri2019empirical,chowdhuri2020revisiting} and stable boundary layers \citep{heisel2023evidence}. Nevertheless, regarding RSL flows, the mixed scale is observed for the first time while looking at the coherence between the $u$ and $w$ signals. Therefore, the observations from Fig. \ref{fig:4} raise two important questions. First, what physical processes are associated with large $\Gamma^2_{u,w}$ values at heights deep within the canopy despite their cospectral amplitudes being nearly zero?; Second, what does a mixed scaling signify in terms of the eddy structures in canopy flows?.

\begin{figure*}[h]
\centering
\includegraphics[width=1\textwidth]{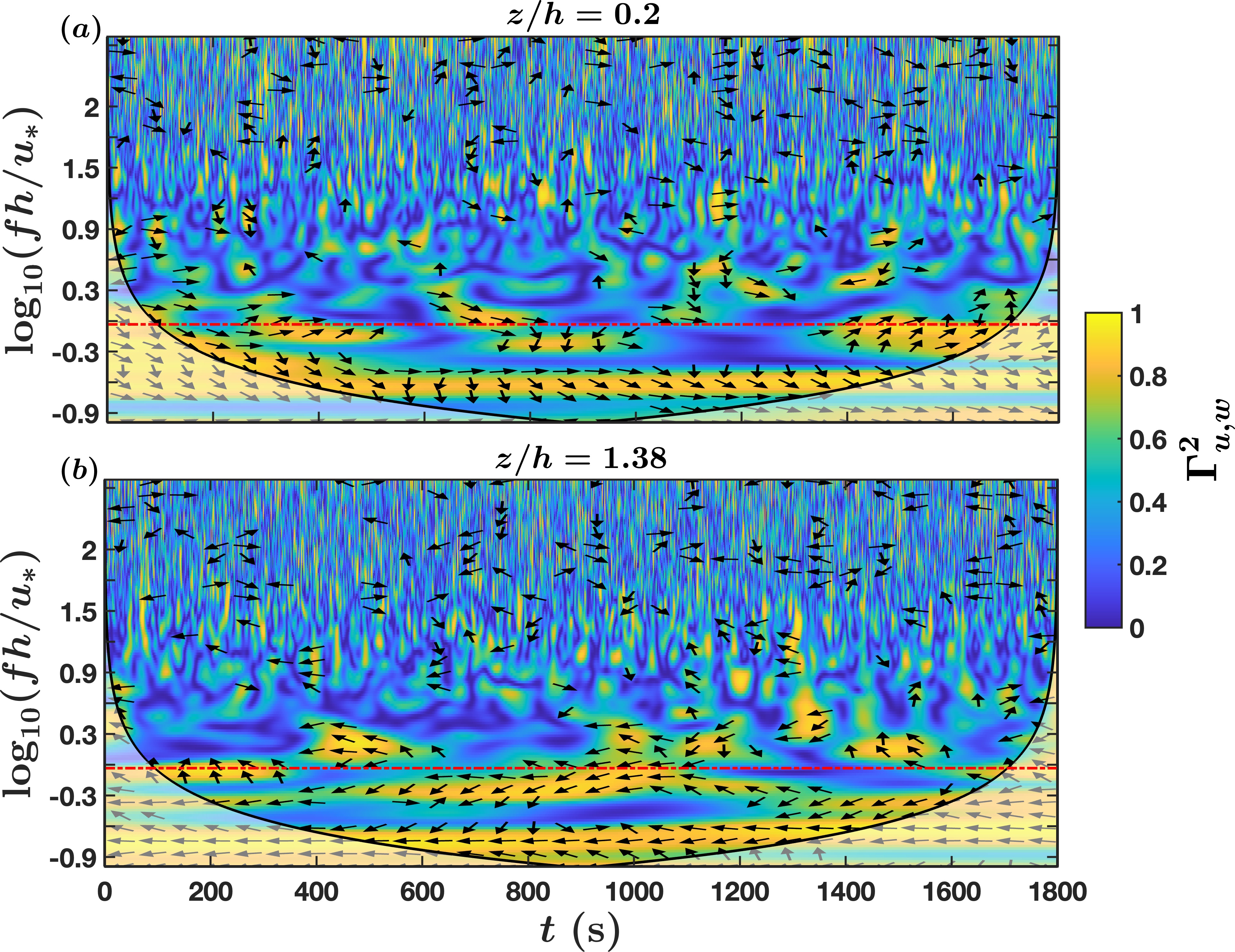}
  \caption{The wavelet cross-scalograms between $u$ and $w$ signals are shown for a specific 30-min run from the GoAmazon dataset, corresponding to the measurement levels (a) $z/h=0.2$ and (b) $z/h=1.38$. The colored contours represent the squared coherence $\Gamma_{u,w}^2$ (see the color bar) while the vertical and horizontal axes denote the scaled frequencies ($\log_{10}(fh/u_{*})$) and time instants ($t$ in seconds), respectively. The arrows show the direction of the momentum transfer --- when they point towards right or towards left it indicates that the $u$ and $w$ signals are either in phase or out of phase. The arrows pointing vertically up or down indicate no transfer of momentum as the phase angles are either $90^{\circ}$ or $270^{\circ}$. The black curved lines represent the cone-of-influence, beyond which the values cannot be trusted. The horizontal red dash-dotted line indicates $fh/u_{*}=1$}
\label{fig:5}
\end{figure*}

To provide a definitive answer to the first question, we investigate the wavelet cross-scalograms between the $u$ and $w$ signals for a specific 30-min run from the GoAmazon dataset (Fig. \ref{fig:5}). These cross-scalograms are computed using the procedure described by \citet{grinsted2004application} and we only show them for two measurement levels, one at a $z/h=0.2$ (Fig. \ref{fig:5}a) and the other at a $z/h=1.38$ (Fig. \ref{fig:5}b). The conclusions remain the same if any other measurement heights or 30-min runs were used in the analysis. In these diagrams, the contours represent the squared coherence $\Gamma^2_{u,w}$, while the horizontal and vertical axes denote the time-instants ($t$ in seconds) and the logarithm values of the scaled frequencies ($\log_{10}(fh/u_{*})$). These frequencies are converted from the wavelet periods and thereafter scaled with the canopy time scale. The arrows depict the phase information between the $u$ and $w$ signals and are shown only for those coherence values that are statistically significant. If the arrows point towards right, then it means the $u$ and $w$ signals are in phase, or in other words, the $uw$ values are positive. Conversely, when the arrows point towards the left, they indicate that the $u$ and $w$ signals are out of phase with the $uw$ values being negative. No momentum is transported when the arrows are vertically up or down since the phase angles in such cases remain at either $90^{\circ}$ or $270^{\circ}$. According to the quadrant nomenclature \citep{wallace2016quadrant}, positive $uw$ values belong to the counter-gradient quadrants (outward- and inward-interaction) while the negative ones belong to the gradient quadrants (ejection and sweep). 

From Fig. \ref{fig:5}a one can see that the large coherence values at larger time scales ($\log_{10}(fh/u_{*})\leq 0$) of the flow are mostly associated with the phase arrows pointing towards the right. Therefore, at within canopy levels, the large values of $\Gamma_{u,w}^2$ are associated with the counter-gradient momentum transport events. The situation reverses for heights above the canopy (Fig. \ref{fig:5}b), where the large coherence levels coincide with the gradient momentum transport (arrows pointing towards the left). The abundance of the counter-gradient momentum events, occurring at heights $z/h < 1$, is also reflected in the PDFs of event lengths, computed separately for each of the four quadrants (see Fig. S3 in the Supplementary Material). 

The PDFs of event lengths, corresponding to the counter-gradient quadrants, clearly display heavy tails at larger time scales for heights $z/h < 1$ (Figs. S3c--d). However, these heavy tails disappear progressively as the heights approach the canopy top. On the other hand, the event length PDFs of the gradient quadrants remain remarkably invariant with height (Figs. S3a--b). Therefore, it is apparent that the coherent structures inside the canopy air space comprise both of counter-gradient and co-gradient events. This information is used to shed light on the mixed time scale. 

\begin{figure*}[h]
\centering
\includegraphics[width=1\textwidth]{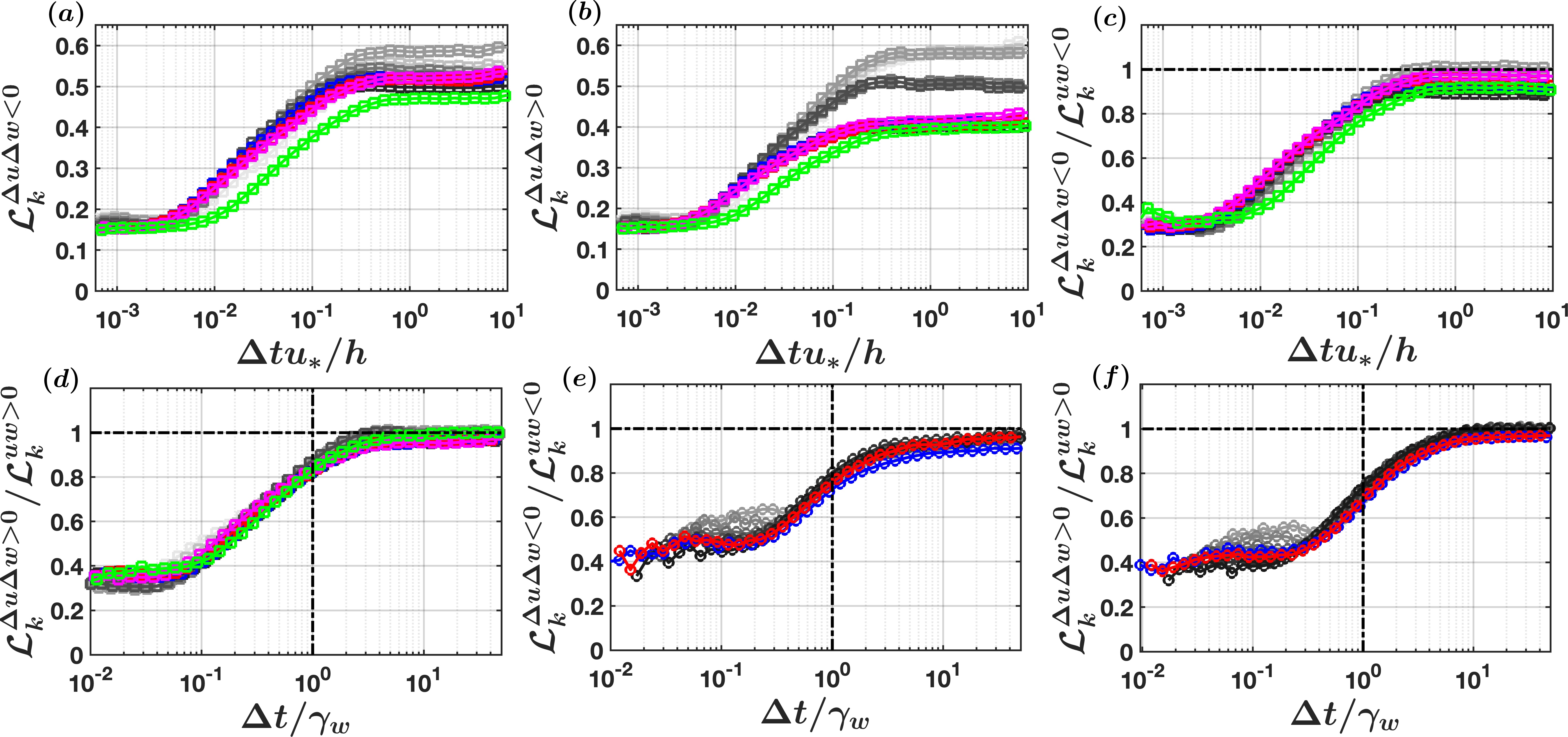}
  \caption{For the RSL datasets, the scale-wise evolution of the $\mathcal{L}_{k}$ values of the event lengths are shown separately for the (a) gradient ($\mathcal{L}_{k}^{\Delta u\Delta w<0}$) and (b) counter-gradient ($\mathcal{L}_{k}^{\Delta u\Delta w>0}$) momentum transport. The time-lags ($\Delta t$) are normalized by the canopy time scale. (c), The $\mathcal{L}_{k}^{\Delta u\Delta w<0}$ values are plotted similarly as in (a) but those are scaled by the full-signal values, i.e. $\mathcal{L}_{k}^{uw<0}$. The $\mathcal{L}_{k}^{uw<0}$ values are obtained after conditionally sampling the negative instantaneous $uw$ signals and computing the $\mathcal{L}_{k}$ values of their time scales. (d) A similar exercise is repeated for the counter-gradient momentum transport but the time-lags in that case are normalized by $\gamma_{w}$. For comparison purposes, in (e,f) the scaled curves are shown from the ASL datasets (SLTEST and CAIPEEX) by separating the momentum transport into its gradient and counter-gradient components. For both (e) and (f), the time-lags are normalized by $\gamma_{w}$.}
\label{fig:6}
\end{figure*}

To accomplish that objective, we first conditionally sample the positive and negative momentum flux events at each scale of the flow. At a time lag of $\Delta t$, the positive momentum flux events are represented as $\Delta u\Delta w>0$ while the negative ones are $\Delta u\Delta w<0$. In Figs. \ref{fig:6}a--b, we plot the $\mathcal{L}_{k}$ values of their event lengths, separately for the negative ($\mathcal{L}_{k}^{\Delta u\Delta w<0}$) and positive ($\mathcal{L}_{k}^{\Delta u\Delta w>0}$) components. Here, the time lags $\Delta t$ are normalized by the canopy time scale. From Fig. \ref{fig:6}a it is conspicuous that the $\mathcal{L}_{k}^{\Delta u\Delta w<0}$ curves do not appreciably change with heights. Notwithstanding that the average momentum fluxes within the canopy are nearly zero (see Fig. \ref{fig:4}a), this finding suggests that the ejection and sweep motions in the canopy sub layers carry the signatures of the mixing layer eddies although they do not actively transport any momentum.

Conversely, a strong height dependence is noted for $\mathcal{L}_{k}^{\Delta u\Delta w>0}$ curves at time scales beyond $\Delta t u_{*}/h=0.05$. In the frequency domain, the time scale $\Delta t u_{*}/h=0.05$ is converted to $fh/u_{*}=20$, which roughly corresponds to the secondary peak in the velocity spectra for heights deep within the canopy (see Fig. S4 in the Supplementary Material). A similar phenomenon is also evident from the coherence spectra in Fig. \ref{fig:4}b, where a dip in the $\Gamma_{u,w}^2$ values are seen at $fh/u_{*} \approx 20$. \citet{poggi2004effect} show that a secondary high-frequency peak in the velocity spectra is associated with the presence of von K\`{a}rm\`{a}n vortices shed by the plant elements, which corresponds well with a scaled frequency value of $fd/\overline{U}=0.21$, where $d$ is the trunk diameter and $\overline{U}$ is the local mean wind speed. The GoAmazon observations show $\overline{U}/u_{*} \approx 0.5$ within canopy and if we assume the ratio between the tree height ($h$) and trunk diameter ($d$) is around 100 for the Amazon forest, then $fd/\overline{U}=0.21$ matches well with $fh/u_{*} \approx 20$.

Beyond $\Delta t u_{*}/h=0.05$, the values of $\mathcal{L}_{k}^{\Delta u\Delta w>0}$ remain significantly higher for the heights within the canopy, with the largest values typically being found at the lowest three levels. Accordingly, the plateaus attained by the $\mathcal{L}_{k}^{\Delta u\Delta w>0}$ curves also show a similar behaviour. Therefore, one is more likely to encounter long lasting counter-gradient events near the forest floor rather than at the higher heights. Statistically speaking, we can thus conclude that within the canopy air space the lengths of the counter-gradient events decrease with height. All these observations from Figs. \ref{fig:6}a--b are consistent with the event length PDFs being presented in Fig. S3. 

The differences between the positive and negative flux events are illustrated quite nicely in Fig. \ref{fig:10} of Appendix \ref{app_C}, where we present both the line and contour plots of the ratio, $\mathcal{L}_{k}^{\Delta u\Delta w<0}/\mathcal{L}_{k}^{\Delta u\Delta w>0}$. Now, the $\mathcal{L}_{k}^{\Delta u\Delta w<0}$ curves can be collapsed even better if the $\mathcal{L}_{k}^{\Delta u\Delta w<0}$ values are scaled by their full signal values, i.e. $\mathcal{L}_{k}^{uw<0}$ (Fig. \ref{fig:6}c). The $\mathcal{L}_{k}^{\Delta u\Delta w>0}$ curves, on the other hand, collapse remarkably well when the time scales are normalized by the integral time scale of the vertical velocity ($\gamma_{w}$) and the $\mathcal{L}_{k}^{\Delta u\Delta w>0}$ values are scaled by $\mathcal{L}_{k}^{uw>0}$ (Fig. \ref{fig:6}d). In fact, in these scaled coordinate systems, the two RSL datasets (GoAmazon and Duke forest) agree quite strongly with each other. The collapse is however poor for the $\mathcal{L}_{k}^{\Delta u\Delta w>0}/\mathcal{L}_{k}^{uw>0}$ curves if the time lags were scaled either by $\gamma_{u}$ or the canopy time scale (see Fig. S5 in the Supplementary Material). The existence of two different time scales to collapse the event lengths of negative and positive momentum flux events is not applicable for atmospheric surface layer flows, since in those cases $\gamma_{w}$ appears to be the only relevant time scale (Figs. \ref{fig:6}e--f). 

In a nutshell, the transport of gradient and counter-gradient momentum inside the canopy air space is accomplished through two different eddy processes. The gradient momentum exchanges occur at a time scale commensurate with the canopy time scale ($h/u_{*}$). However, for counter-gradient momentum exchanges, $\gamma_{w}$ emerges to be the suitable scale. Since the two exchange processes coexist, an interaction between these two scales gives rise to the mixed time scale as observed in Fig. \ref{fig:4}c. We next present a conceptual model to explain this in terms of the canopy flow physics. 

\subsection{A conceptual model of RSL flows}
\label{conceptual model}
\begin{figure*}[h]
\centering
\includegraphics[width=1\textwidth]{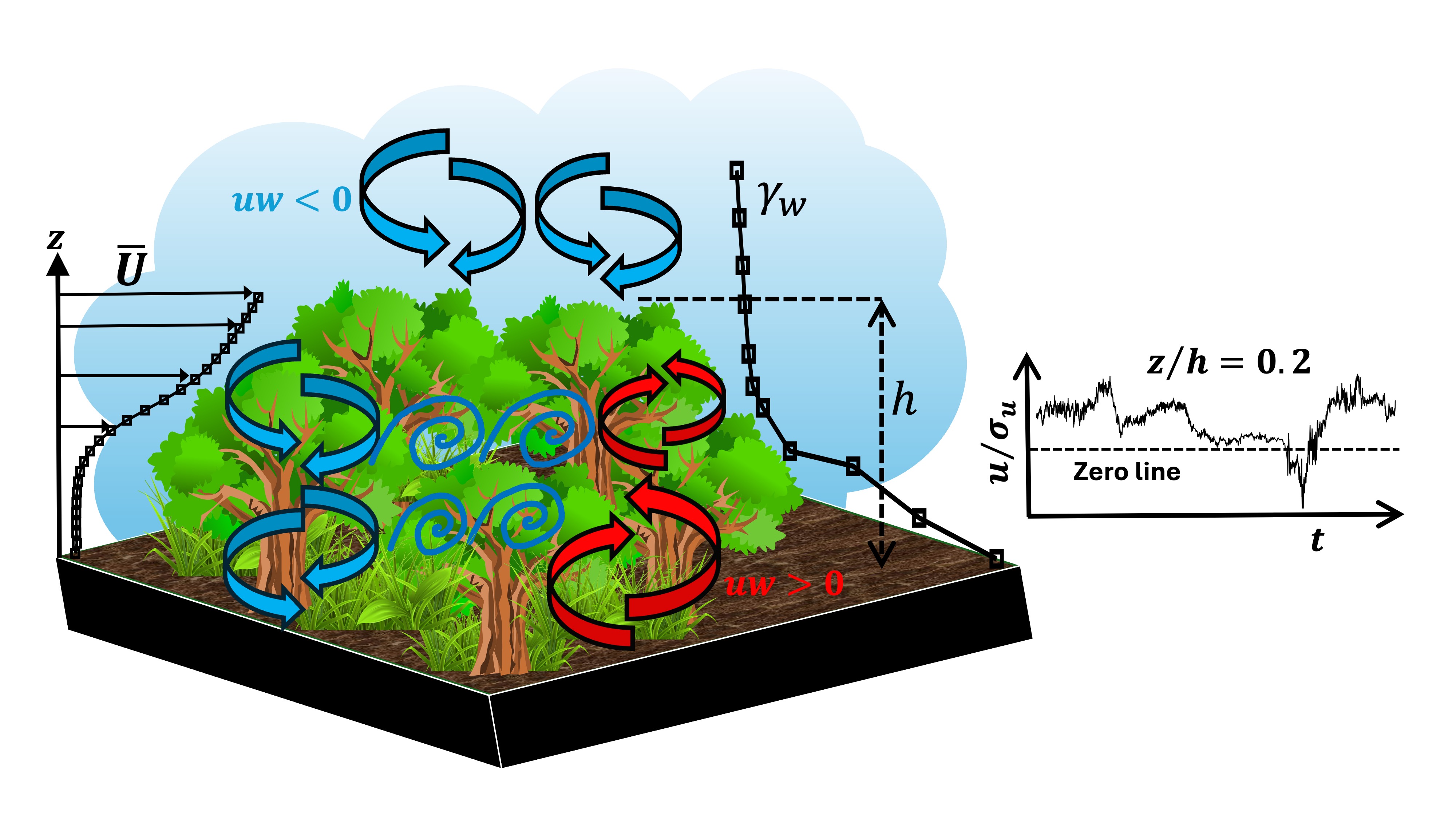}
  \caption{A schematic diagram to explain the nature of momentum transport in canopy flows. The canopy-scale eddies (blue curved arrows, $uw<0$) progressively penetrate down within the canopy and as they do, due to the presence of the obstacles (the tree trunks and leaves), they shed more and more smaller-scale eddies (light blue spiral arrows), possibly the von K\`{a}rm\`{a}n vortices. These vortices merge together and form the counter-gradient eddies (red curved arrows, $uw>0$), whose sizes are the largest when closer to the forest floor. The preponderance of these counter-gradient eddies near the forest floor causes the integral time scales of the vertical velocity ($\gamma_{w}$) to increase as the height ($z$) decreases. Within the canopy air space, although the mean wind speed ($\overline{U}$) approaches zero, the interaction between the gradient and counter-gradient eddies cause sudden bursts and lull periods in the instantaneous wind fluctuations ($u$), thereby contributing to its intermittency. This is illustrated through a segment of $u/\sigma_{u}$ time-series, collected at a height of $z/h=0.2$.}
\label{fig:7}
\end{figure*}

In Fig. \ref{fig:7} we present a schematic diagram of a conceptual model that summarizes the findings presented so far. As observed in Fig. \ref{fig:6}, the canopy time scales govern the negative momentum flux events and therefore the origin of these events is related to the turbulent processes occurring at the canopy top. Accordingly, the eddies generated through the Kelvin-Helmholtz (KH) instability at the canopy top carry negative momentum along with them (shown as blue curved arrows). As these KH eddies penetrate deep within the canopy, due to the presence of the obstacles (tree trunks and foliage), they shed more and more smaller-scale eddies (shown as light blue spiral arrows), possibly the von K\`{a}rm\`{a}n vortices. Our results in Fig. \ref{fig:6}b suggest that the $\mathcal{L}_{k}^{\Delta u\Delta w>0}$ curves separate with height at scales beyond the von K\`{a}rm\`{a}n scale ($\Delta t u_{*}/h \geq 0.05$), and they keep on increasing until a plateau is reached. A plausible interpretation is that the smaller scale von K\`{a}rm\`{a}n eddies progressively merge together and give rise to the large counter-gradient momentum events. Since one would expect a preponderance of these smaller scale eddies as the heights approach the forest floor (i.e. the wall), large counter-gradient events are possible due to an increased possibility of these eddies to coalesce. This is illustrated in Fig. \ref{fig:7} by curved red arrows that carry positive momentum flux and their sizes decrease with height within the canopy. 

Furthermore, the time scales of these counter-gradient events are governed by $\gamma_{w}$ (Fig. \ref{fig:6}d), and previous evidences suggest that $\gamma_{w}$ decreases with height in canopy sub layers \citep{launiainen2007vertical,chamecki2013persistence}. The same is true for the GoAmazon dataset, represented as a black line with square markers in Fig. \ref{fig:7}. This behaviour of $\gamma_{w}$ is in stark contrast with ASL flows, where the $\gamma_{w}$ values are supposed to increase with height as a consequence of the attached-eddy hypothesis. We highlight this difference in Fig. S6 of the Supplementary Material, where we compare the vertical profiles of $\gamma_{w}$ between the GoAmazon and SLTEST datasets. Regarding canopy flows, from our results one can infer that the increase in $\gamma_{w}$ with decreasing height is related to the presence of large counter-gradient eddies at the forest floor. The coexistence of these gradient and counter-gradient eddies make the turbulent wind field inside a canopy strongly intermittent. Although over time the negative and positive momentum events cancel each other, their instantaneous contributions appear to be the root cause behind intermittency. 

For example, the negative flux events bring excess momentum at the canopy top to the heights within the canopy and therefore contribute to the sudden gusts in the subcanopy wind speed. On the other hand, when the flux events are positive they cause the air to lose its momentum to the layer above, which eventually manifests as sudden lull periods in the wind speed. As a result, the turbulent wind inside a canopy displays strong intermittent features, such as sudden increases followed by a decrease. This occurs even though the mean wind speed ($\overline{U}$) approaches zero as the forest floor is reached. This is demonstrated in Fig. \ref{fig:7} through a sample $u$ time series from the GoAmazon dataset at a height of $z/h=0.2$. 

This conceptual model rests on the hypothesis that the long-duration events are formed through a merging process with smaller-scale structures coalescing together. It remains to be seen whether any alternate hypothesis linked to the large ABL-scale motions can also explain the origin of these counter-gradient events. This exploration is beyond the scope of this present study. We present our conclusions and future outlooks in the next section.

\section{Conclusion}
\label{conclusion}
In this study, we developed an $\mathcal{L}$-moment based event framework to model the dynamics of a homogeneous canopy flow. Primarily, this study attempts to answer a few fundamental issues in canopy turbulence, which are: up to what depth do the canopy-scale eddies penetrate, what causes $\gamma_{w}$ to decrease with height, and how exactly these two scales interact to shape up the momentum transport in canopy flows? 

To answer these questions, we considered the turbulent fluctuations at each scale of the flow to consist of an alternating positive and negative event chronicle with certain sizes. The PDFs of these event lengths become progressively heavy-tailed as the scales increase due to the presence of coherent structures in turbulent flows. The conventional statistical moments cannot quantify these heavy-tailed distributions, since those estimates do not converge satisfactorily at all. In that respect, the $\mathcal{L}$-moments are more useful, as these moments were specifically developed to deal with heavy-tailed distributions. Accordingly, the statistical distribution of these event lengths could be accounted for by computing their $\mathcal{L}$-kurtosis values, which quantify the heaviness of their tails. Therefore, the $\mathcal{L}$-moments allow us to reliably represent the event-length distributions, whose PDFs have non-trivial shapes with an extended power-law regime. Such metrics suitable for analysing heavy-tailed distributions could be useful also for wind-energy applications where varying periods of constant wind speeds are known to affect the performance of the wind turbines. The insights obtained from this framework are complemented with a wavelet analysis. 

Our results indicate that the dynamics of the vertical velocity fluctuations encode the effects of canopy-scale eddies in RSL flows. Contrary to some previous literature, we propose that these canopy-scale eddies do penetrate deep within the canopy and are not restricted up to any specific height. However, these eddies become increasingly inactive in terms of downward momentum transport as they penetrate through the canopy. The wavelet analysis reveals that the momentum transport in canopy flows is controlled by a mixed time scale, represented as, ${(\gamma_{w}h/u_{*})}^{1/2}$. Therefore, this time scale is a geometric mean of $\gamma_{w}$ and the canopy time scale. Through further analysis we convincingly demonstrate that the origin of this mixed scale is intimately linked to an interaction between two different eddy processes that transport momentum in the gradient and counter-gradient directions, respectively. This is perhaps the first time where a physical interpretation of mixed time scale is provided. 

Backed by sufficient evidence, we show that the decrease of $\gamma_{w}$ with height is a consequence of the fact that the lengths of the counter-gradient events decrease with height. We hypothesize that these counter-gradient events are formed through a merging process. Our findings suggest that as the eddies at the canopy top penetrate down, due to the presence of the obstacles (tree trunks and foliage), they shed more and more smaller-scale eddies, possibly the von K\`{a}rm\`{a}n vortices. These smaller-scale eddies coalesce and form the counter-gradient eddies, whose sizes are the largest when closer to the forest floor. This proposed mechanism explains the scale-interaction between the canopy-scale eddies and the eddies whose sizes are comparable to the integral scales of vertical velocity. At present, it is not entirely clear how exactly the sign of the momentum transport reverses as the smaller-scale eddies merge together. For future studies, advanced simulations, capable of resolving these smaller scale eddies, are needed to understand this better. 

It is also shown that the coexistence of these gradient and counter-gradient eddies make the turbulent wind field inside a canopy strongly intermittent. Large-eddy simulations (LES) of canopy flows might not capture these large counter-gradient events, since the origin of these events is supposedly tied to the smaller scale eddies that are not resolved by the LES. This likely has an impact on the simulated wind speed inside a canopy, which might appear to be smoother than the real observations. This is indeed an important caveat since a strong intermittent wind field inside the canopy air space has profound implications towards modelling the fire spread through a forest or the transportation of gases or bio-aerosol particles emitted from the forest to the upper atmosphere. Therefore, our results provide a benchmark to test the next generation LES models of canopy flows that capture the effects of these counter-gradient events and thereby intermittency. 

An another future direction is to investigate the role of heterogeneity (e.g., sparse canopies and canopy edge flows) and atmospheric stability on the scales of momentum transport and scalars, such as heat and moisture. Such alterations to canopy structure are often caused by forest management practises, such as clearcutting or thinning. What remains unclear is how cutting down the trees, thereby reducing the leaf area density, would affect the interaction between the canopy- and $\gamma_{w}$-scale eddies. This has far-reaching implications towards designing mitigation strategies to constrain the spread of wildfires through a forest.

\begin{acknowledgements} 
SC and OP thank the support from the Research Council of Finland (grant no. 354298). SC acknowledges the discussion with Jasper Vrugt and Efi Foufoula-Georgio, regarding the application of $\mathcal{L}$-moments in statistics. SC acknowledges Thara Prabhakaran and Anand Karipot for letting him use the CAIPEEX data for this research. SC thanks Keith McNaughton, Khaled Ghannam, Marcelo Chamecki, and Gaby Katul for the use of the SLTEST, GoAmazon, and Duke Forest datasets.
\end{acknowledgements} 

\appendix
\section{Normalized $\mathcal{L}_{k}$ curves}
\label{app_A}
\begin{figure*}[h]
\centering
\includegraphics[width=1\textwidth]{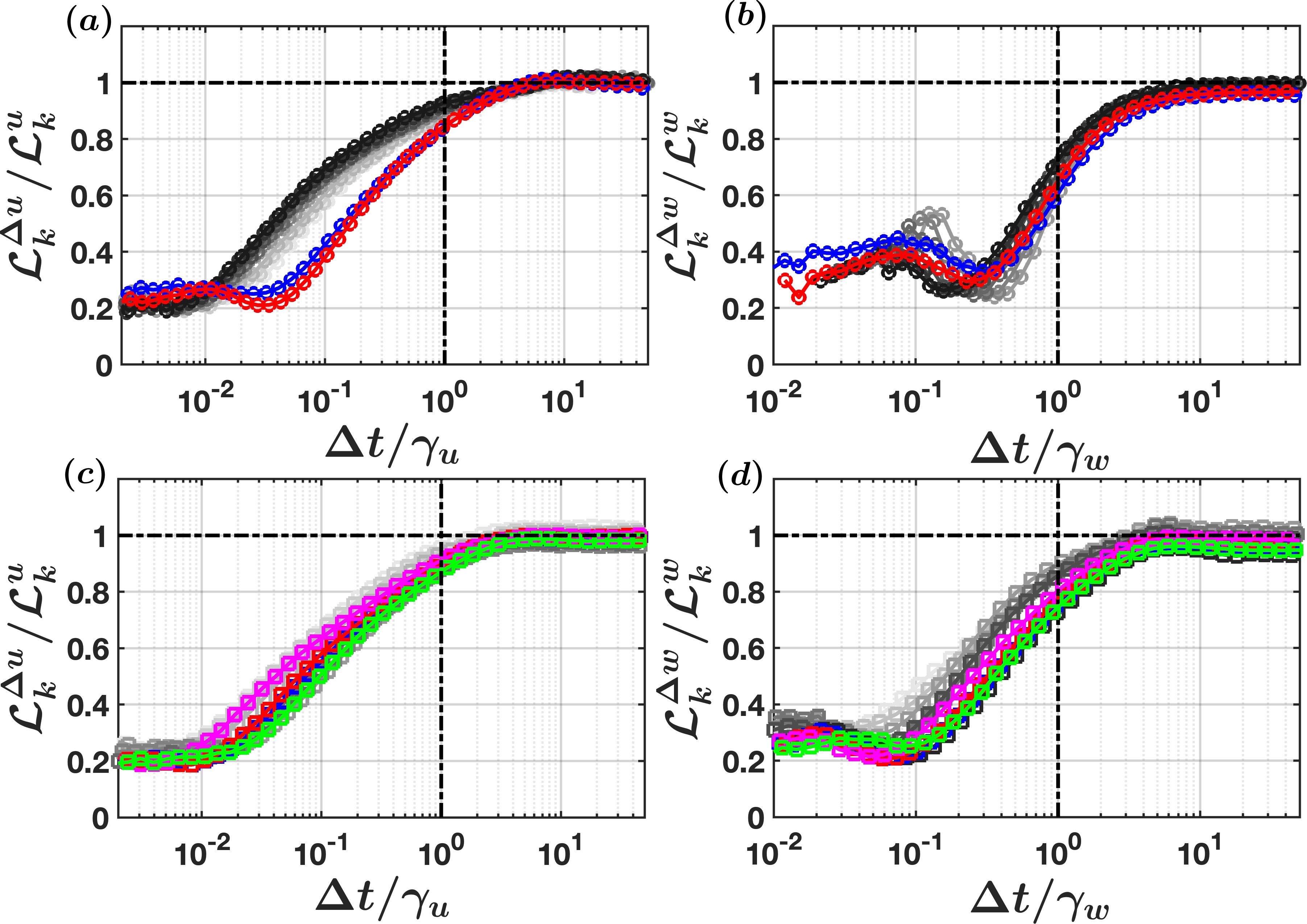}
  \caption{Same as in Fig. \ref{fig:2}, but the scale-wise $\mathcal{L}_{k}$ values ($\mathcal{L}^{\Delta x}_{k}$) are normalized by their respective full-signal values.}
\label{fig:8}
\end{figure*}

In this appendix, we show the scale-wise evolution of the $\mathcal{L}^{\Delta x}_{k}$ ($x=u,w$) values, corresponding to the event lengths of the $\Delta x$ signals at a prescribed time lag $\Delta t$. As opposed to Fig. \ref{fig:2}, instead of presenting the $\mathcal{L}^{\Delta x}_{k}$ values alone, we divide those with the $\mathcal{L}_{k}$ values obtained from the event lengths of the full signal $x$ ($\mathcal{L}^{x}_{k}$). Under such normalization, one can notice from Fig. \ref{fig:8} that as the time lags increase, the $\mathcal{L}^{\Delta x}_{k}$ values approach $\mathcal{L}^{x}_{k}$, regardless of the signal or the flow types. 

Notably, only for the ASL flows, there exist clear differences between the SLTEST and CAIPEEX datasets when the $u$ signals are considered (Fig. \ref{fig:8}a), a feature not immediately apparent from Fig. \ref{fig:2}a. The zero-crossing properties of the $u$ signals are influenced by the presence of the large-scale structures in the flow (Fig. \ref{fig:3}a). Since these large-scale structures are sensitive to the boundary conditions, we attribute the differences in $\mathcal{L}^{\Delta u}_{k}/\mathcal{L}^{u}_{k}$ curves to the varying surface conditions at the SLTEST and CAIPEEX sites. However, for the $w$ signal, we do not observe any such discrepancies as the curves collapse reasonably well among different datasets (Fig. \ref{fig:8}b). 

Nevertheless, the $\mathcal{L}^{\Delta w}_{k}/\mathcal{L}^{w}_{k}$ curves attain their plateaus at scales significantly larger $\gamma_{w}$ (Fig. \ref{fig:8}b). This is more obvious for the ASL flows rather than for the RSL (Fig. \ref{fig:8}b and d). As discussed in Section \ref{contrast}, this difference is intimately linked to how the dynamics of the vertical velocity fluctuations encode the effects of canopy-scale eddies in RSL flows, a mechanism not applicable for ASL flows. 

\section{Third- and fourth-order $\mathcal{L}$ moments}
\label{app_B}
\begin{figure*}[h]
\centering
\includegraphics[width=1\textwidth]{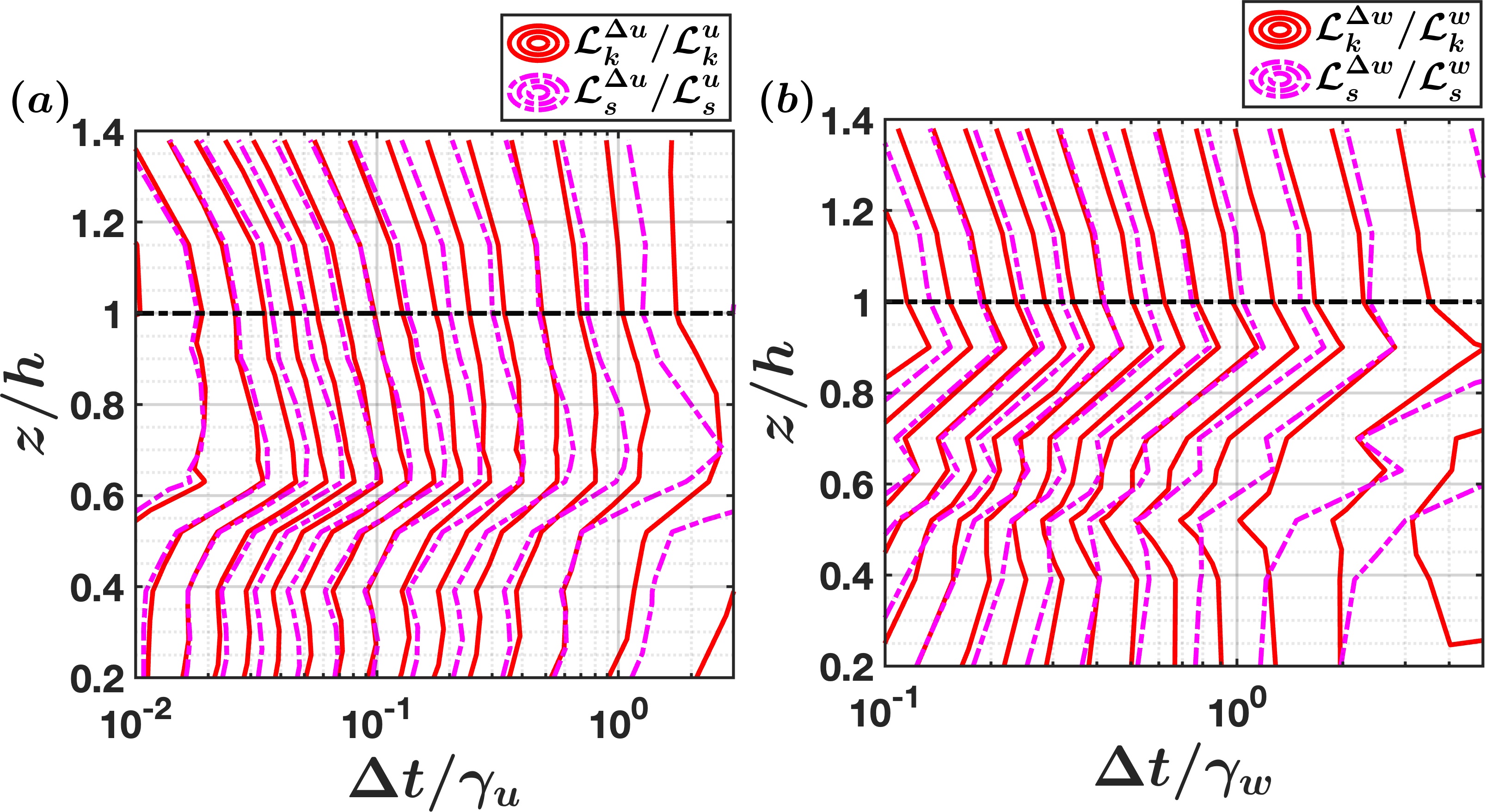}
  \caption{For the GoAmazon dataset, the time-height contour plots of (a) $\mathcal{L}_{k}^{\Delta u}$ and (b) $\mathcal{L}_{k}^{\Delta w}$ are shown as red solid lines. On these contours, the $\mathcal{L}_{s}^{\Delta x}$ ($x=u,w$) values are overlaid (shown as pink dash-dotted lines), which represent the $\mathcal{L}$-skewness moments corresponding to the event lengths of the $\Delta u$ and $\Delta w$ signals. Both $\mathcal{L}$-skewness and -kurtosis moments are scaled by their respective full-signal values.}
\label{fig:9}
\end{figure*}

From a statistical perspective, the heavy tails of a PDF can also be quantified through skewness instead of kurtosis. In that respect, we compare the time-height contour plots of both $\mathcal{L}$-kurtosis and $\mathcal{L}$-skewness moments for the canopy flows. Figure \ref{fig:9} shows these contour plots separately for the $u$ and $w$ signals from the GoAmazon dataset. The contours here represent the normalized $\mathcal{L}$-kurtosis (shown as thick red lines) and $\mathcal{L}$-skewness moments (pink dash-dotted lines) of the event lengths, corresponding to the $\Delta u$ and $\Delta w$ signals, respectively. The time lags are scaled by the integral scales, while the heights are normalized by the canopy height. No appreciable difference can be noted between the two contours, thereby implying that the $\mathcal{L}$-kurtosis or $\mathcal{L}$-skewness can be used interchangeably to address the heavy tails of the event length PDFs at any desired scale of the flow. In general, this finding reiterates the robustness of using $\mathcal{L}$-moments to characterize the heavy-tailed PDFs of any stochastic signal. 

\section{Ratios of $\mathcal{L}_{k}$ values}
\label{app_C}

\begin{figure*}[h]
\centering
\includegraphics[width=1\textwidth]{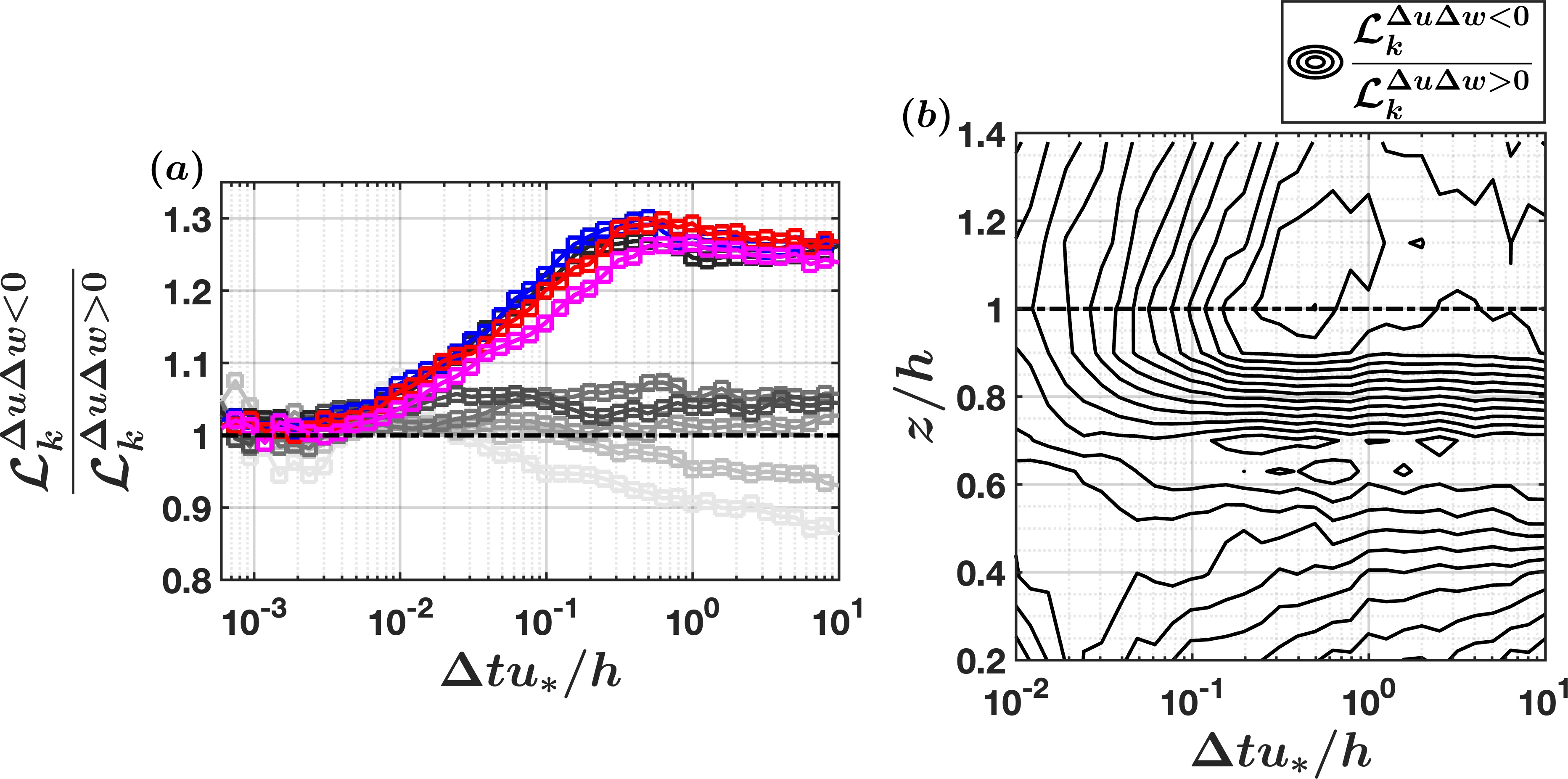}
  \caption{For the GoAmazon dataset, the ratios between $\mathcal{L}_{k}^{\Delta u\Delta w<0}$ and $\mathcal{L}_{k}^{\Delta u\Delta w>0}$ values are shown as (a) line, and (b) time-height contour plots. The time-lags are normalized by the canopy time scale while the heights are scaled by the canopy height.}
\label{fig:10}
\end{figure*}

In this appendix, we demonstrate how the differences in the event-length statistics of gradient and counter-gradient momentum fluxes delineate between the eddy motions occurring at within and above the canopy air space. For that purpose, in Fig. \ref{fig:10}a, we plot the ratios of $\mathcal{L}_{k}$ values computed separately for the event lengths of $\Delta u\Delta w<0$ and $\Delta u\Delta w>0$ signals. One can clearly see, the ratio, $\mathcal{L}_{k}^{\Delta u\Delta w<0}/\mathcal{L}_{k}^{\Delta u\Delta w>0}$, remains closer or smaller than unity for heights within the canopy at all scales of the flow. However, for heights just above and beyond the canopy, these ratios exceed unity considerably. Accordingly, this indicates the importance of the counter-gradient eddies prevailing within the canopy air space. 

Such differences between the within and above canopy turbulence can also be visually identified if one plots the contours of $\mathcal{L}_{k}^{\Delta u\Delta w<0}/\mathcal{L}_{k}^{\Delta u\Delta w>0}$ against the normalized time-lags and heights (Fig. \ref{fig:10}b). By seeing how these contours orient themselves, one can notice the existence of two distinct zones separated at a height of $z/h \approx 0.9$, i.e. very close to the canopy top. Therefore, the findings from Fig. \ref{fig:10} lend credence to our conceptual model as discussed in Section \ref{conceptual model}. 

\bibliographystyle{spbasic_updated}     
\bibliography{sample_library} 
\end{document}